\documentclass[apj,numberedappendix,twocolappendix]{emulateapj}
\usepackage{apjfonts}

\usepackage{graphicx}
\usepackage{mathrsfs}
\usepackage{natbib}
\bibpunct[; ]{(}{)}{;}{a}{}{,}

\def\totd{{\mathrm{d}}}
\newcommand{\be}{\begin{equation}}
\newcommand{\ee}{\end{equation}}
\defcitealias{M12}{M12}

\shorttitle{NUCLEAR DOMINATED ACCRETION FLOWS IN 2D}
\shortauthors{FERN\'ANDEZ \& METZGER}

\begin{document}

\title{Nuclear Dominated Accretion Flows in Two Dimensions.\\
I. Torus evolution with Parametric Microphysics}
\author{Rodrigo Fern\'andez\altaffilmark{1,2}}
\author{Brian D. Metzger\altaffilmark{2,3}}
\altaffiltext{1}{Institute for Advanced Study. Einstein Drive, Princeton, NJ 08540, USA.}
\altaffiltext{2}{Einstein Fellow}
\altaffiltext{3}{Department of Astrophysical Sciences, Peyton Hall, Princeton University, Princeton, NJ 08544, USA.}

\begin{abstract}
We explore the evolution of radiatively inefficient accretion disks in which
nuclear reactions are dynamically important (`Nuclear Dominated Accretion
Flows', or NuDAFs).  Examples of such disks are those generated by the merger of a white dwarf
with a neutron star or black hole, or by the collapse of a rotating
star.  Here we present two-dimensional hydrodynamic simulations
that systematically explore the effect of adding a single nuclear reaction to
a viscous torus.  The equation of state, anomalous shear
stress, and nuclear reactions are given parametric forms.  
Our results point to the existence of two qualitatively different
regimes of NuDAF evolution: (1) steady accretion with quiescent
burning; or (2) detonation of the disk.  
These outcomes are controlled primarily by
the ratio $\Psi$ of the nuclear energy released to the
enthalpy at the burning radius. Disks detonate if
$\Psi$ exceeds a critical value $\Psi_{\rm crit} \sim 1$, and if burning
occurs in regions where neutrino cooling is unimportant.  
Thermonuclear runaways are seeded by the turbulent mixing of hot
ash with cold fuel 
at the burning front. Disks with $\Psi < \Psi_{\rm crit}$ do not explode, 
but instead power a persistent collimated outflow of
unbound material composed primarily of ash, with a mass-loss rate that increases
with $\Psi$.  
We discuss the implications of our results for 
supernova-like counterparts 
from astrophysical events in the NuDAF regime.  
In particular, detonations following a white dwarf - neutron star merger
could account for some subluminous Type Ia supernovae, such as the class
defined by SN 2002cx.  
\end{abstract}

\keywords{accretion, accretion disks --- nuclear reactions, nucleosynthesis, abundances 
	  --- white dwarfs --- hydrodynamics --- stars: winds, outflows 
          --- supernovae: general}

\maketitle

\section{Introduction}

Many violent and visually spectacular astrophysical events involve stellar
material collapsing at high rates onto a central compact object.  Examples
include the core collapse of a massive star to form a neutron star (NS) or
black hole (BH) remnant \citep{Woosley+02}; the accretion-induced collapse of a
white dwarf (WD) \citep{Nomoto&Kondo91}; the merger of binary compact objects,
including various combinations of WDs, NSs, and BHs; the binary merger of a
Helium star with a NS or BH (e.g.,~\citealt{Fryer&Woosley98}); and the inspiral
and tidal disruption of a planet that is engulfed by the envelope of its host
star (e.g.,~\citealt{Siess&Livio99}).  Although only some of these events have
yet been unambiguously observed, this situation may change soon as the result
of new transient surveys coming online across the electromagnetic spectrum.

In many of the above examples, angular momentum places a key dynamical role,
such that a significant fraction of the infalling matter forms a
rotationally-supported torus around the central object.  If the accretion rate
through such a disk is sufficiently high, then the heat generated is trapped in
the flow instead of being efficiently radiated away (e.g.,
\citealt{Chevalier93}).  Such disks are examples of what is generally termed a
\emph{Radiatively Inefficient Accretion Flow} (RIAF).\footnote{The high density
RIAFs discussed here are to be distinguished from those that occur at much
lower densities due to inefficient two-body collisional cooling, and which are
used to model low-luminosity AGN.}

Previous theoretical work on RIAFs has focused on understanding their spatial
structure and on determining the processes that control the radial transport of mass,
energy, and angular momentum through the disk. The ADAF model \citep{Narayan&Yi94,Narayan&Yi95}
assumes that angular momentum is transported outward by turbulent stresses (mediated by,
e.g., the magnetorotational instability [MRI]; \citealt{Balbus&Hawley98})
and that thermal energy is advected through the inner boundary of the disk at
small radii. The CDAF model \citep{Quataert&Gruzinov00} postulates that outward
angular momentum transport is offset by \emph{inward} transport by
convection. The resulting accretion rate is much lower than that arising from
ADAFs, though all of the disk material is eventually accreted by the central object.
In contrast, the ADIOS model \citep{Blandford&Begelman99} assumes that
most of the accretion energy powers an unbound outflow from the disk, such that
the majority of the material never reaches the central object.

Numerous two- and three dimensional global simulations of RIAFs have been performed  in
hydrodynamics (2D -  \citealt{stone1999},
\citealt{Igumenshchev&Abramowicz00}; 3D - \citealt{Igumenshchev+00}) as well as
in magnetohydrodynamics [MHD] (2D - \citealt{stone2001}; 3D -
\citealt{hawley2001}, \citealt{machida2001}, \citealt{Hawley&Balbus02}, \citealt{igumenshchev2003}, 
\citealt{pen2003}, \citealt{pang2011}, \citealt{mckinney2012}, \citealt{narayan2012}).
A key conclusion of most such simulations to date is that the net accretion
rate through the inner edge of the torus at $r = r_{\rm in}$ is substantially
reduced (by a factor $\sim r_{\rm in}/R_{0}$) from the feeding rate set by the
outer torus at $r = R_{0}$, consistent with both CDAF and ADIOS models 
(e.g., \citealt{yuan2012}).
However, the relative importance of truly unbound outflows versus simply
large-scale bound convective motions remains under debate
(e.g.,~\citealt{Abramowicz+00}).

Here we focus on another key feature of RIAFs in collapsar and merging compact
object scenarios: the nuclear composition of the accreting material and its
effect on the dynamics of the accretion disk.  If the potential well of the
primary has an appropriate depth, then the inflowing matter becomes
sufficiently hot and dense for nuclear reactions to become dynamically
important and to generate radioactive material.  Depending on how nuclear
burning affects the accretion rate onto the central object, or the properties
of outflows from the disk, this can greatly impact predictions for the
observational signatures of these events.

\citet[hereafter M12]{M12}
constructed a steady-state, height-integrated
model of accretion following the tidal disruption of a WD by a BH or NS,
including the effects of nuclear burning on both the thermodynamics and
composition of the accreting material.  He adopted an ADIOS-type model, in
which the properties of the outflow from each radius in the disk are calculated
by locally balancing wind cooling with viscous and nuclear heating.  In this
model, the additional heating from nuclear burning acts to enhance the mass
loss rate in the outflow over its nominal value without nuclear burning.  \citetalias{M12}
showed that at radii $r \lesssim 10^{9}$ cm, the midplane density and
temperature become sufficiently high to burn the initial white dwarf material
into increasingly heavier elements (e.g., Mg, Si, S, Ca, Fe, and Ni) at
sequentially smaller radii.  The disk structure thus resembles the onion-like
structure of a massive star, but one which evolves on a much more rapid
timescale.  The combined outflow from all annuli in the disk is composed
primarily of unburnt WD material from the outer disk, along with a smaller
fraction of intermediate mass elements and radioactive $^{56}$Ni originating from smaller radii. 

In some regions of the disk \citetalias{M12} found that the energy released by nuclear
reactions is at least comparable to that released viscously, thus motivating
him to term this novel accretion regime a \emph{Nuclear Dominated Accretion
Flow} (NuDAF).  Because RIAFs are already only marginally bound, even a modest
additional source of energy can have a significant impact on its structure.  In
other words, the properties of NuDAFs could in principle differ markedly from
the predictions of normal ADAF/CDAF/ADIOS models.

Given its many simplifying assumptions, the \citetalias{M12} model cannot address several
key questions relevant to NuDAFs.  These include (1) the true effects of
nuclear burning on the quantity and composition of disk outflows; (2)
multi-dimensional effects, such as radial convective mixing or the diffusion of
burnt fuel upstream; (3) the long-term, global evolution of the disk; and (4)
the thermal stability of the inflow, which is subtle due to the sensitive
temperature dependence of the nuclear reaction rates.  \citetalias{M12} found that thermal
stability depends critically on what is assumed about the pressure dependence
of wind cooling.  If the disk becomes thermally unstable, possible outcomes
range from a complex `limit cycle' behavior (e.g.,~periods of inflow followed by
intense rapid burning) to a complete dynamical explosion.

In this series of papers we explore the effect of nuclear burning on the
structure and evolution of RIAFs by means of axisymmetric hydrodynamic
simulations.  In paper I (this work) we explore the general dynamical
properties of NuDAFs and their outflows, and compare them to the known case
where nuclear burning is absent (e.g., \citealt{stone1999}).  As in \citetalias{M12}, our
study focuses on disks created by the tidal disruption of a WD by a NS or BH
binary companion, in part because the global properties of the torus are
relatively well-constrained by the parameters of the binary.  However, many of
the conclusions we reach should apply to other similar astrophysical events,
such as the collapse of a rotating star.

In this initial study we make several simplifications in order to clearly
identify the processes controlling the dynamics, and to allow an efficient
exploration of parameter space.  The most important of these approximations is
to parameterize the angular momentum transport in the disk by an anomalous
shear viscosity.  We also adopt an ideal gas equation of state with a single
adiabatic index, and for ease include only one nuclear reaction.  In a
subsequent paper we will use a realistic equation of state (EOS) and extend our
nuclear reaction chain to a full $\alpha$-reaction network, thus enabling a
more reliable prediction of observational signatures.

The paper is organized as follows.  In \S\ref{s:model} we describe the physics
included in our model.  Details on the numerical implementation are provided in
\S\ref{s:numerical}.  Results are presented in \S\ref{s:results}, beginning
with an overview of a fiducial model, followed by sub-sections on
exploding and quiescent models.
Implications from our results are discussed in \S\ref{s:discussion}, and
a bulleted summary of our conclusions is given in \S\ref{s:summary}.  Conversion
of physical to code units and reaction rates employed are described in
Appendix~\ref{s:units_reactions}.  Appendix~\ref{s:code_tests} documents tests
of our numerical code.  An analytic model for the conditions leading to disk
detonation via the Zel'dovich mechanism is provided in Appendix
\ref{s:zeldovich}.

Readers uninterested in technical details are encouraged to skip directly to 
\S\ref{s:overview}.

\section{Physical Model}
\label{s:model}

As a clear physical testbed to study NuDAFs, 
we focus on disks that form when a WD is tidally disrupted by a companion NS or BH in 
a close binary system.
We first review the characteristic properties of 
these disks, and then discuss the physical approximations made in this study. 

\subsection{Disk Properties}

As discussed in \citet{fryer1999} and \citetalias{M12}, whether the WD is disrupted by the
primary depends on whether mass transfer is stable or unstable following the
onset of Roche lobe overflow. Stability depends on the mass ratio of the binary
$q = M_{\rm WD}/M_{\rm c}$, with higher values of $q \gtrsim 0.2-0.5$ favoring
disruption (e.g.,~\citealt{Paschalidis+09}). Here $M_{\rm WD}$ and $M_{\rm c}$
are the mass of the WD and the central compact object (NS or BH), respectively.
Once the WD is disrupted, conservation of angular momentum implies that the
material will circularize around the NS/BH at a characteristic radius
\begin{equation}
\label{eq:r_circ}
R_0 = \frac{a_{\rm RLOF}}{1+q},
\end{equation}
where 
\begin{equation}
a_{\rm RLOF} = R_{\rm WD}\frac{0.6q^{2/3} +\ln(1+q^{1/3})}{0.49q^{2/3}}
\end{equation}
is the orbital separation at Roche lobe overflow \citep{eggleton1983} and 
$R_{\rm WD}$ is the WD radius.  Note that equation~(\ref{eq:r_circ}) neglects 
the self-gravity of the disk.  The WD radius $R_{\rm WD}$ is related to its 
mass by \citep{nauenberg1972,fryer1999}
\begin{equation}
\label{eq:mass_radius}
R_{\rm WD} \simeq 10^9 \left(\frac{M_{\rm WD}}{0.7M_\sun}\right)^{-1/3}
             \left[ 1 - \left(\frac{M_{\rm WD}}{M_{\rm CH}}\right)^{4/3}\right]^{1/2}
             \left(\frac{\mu_e}{2} \right)^{-5/3}\textrm{ cm},
\end{equation}
where $M_{\rm CH}\simeq 1.45 (\mu_e/2)^{-2}\,M_\sun $ is the Chandrasekhar mass
and $\mu_e$ is the mean molecular weight per electron.  Table~\ref{t:models}
gives the circularization radii for models considered in this paper.

As matter circularizes around the NS/BH, a fraction of the kinetic energy of rotation is 
thermalized.  This results in a thick torus (e.g., \citealt{fryer1999}) with a 
scale-height $H_{0} \sim R_{0}/2$ and an average density
\begin{equation}
\label{eq:rho_ave}
\bar \rho \sim \frac{M_{\rm WD}}{R_{0}^{3}} = 1.5\times 10^5 \left( \frac{M_{\rm WD}}{0.6M_\sun}\right)
\left( \frac{10^{9.3}\textrm{ cm}}{R_0}\right)^3\textrm{ g cm}^{-3}.
\end{equation}
The orbital time at the circularization radius is
\begin{equation}
\label{eq:tdyn}
t_{\rm orb} \simeq 40\left(\frac{R_0}{10^{9.3}\textrm{ cm}}\right)^{3/2}
\left( \frac{M_{\rm c}}{1.4M_\sun}\right)^{-1/2}\textrm{ s}.
\end{equation}
Assuming that the internal energy $e_{\rm int}$ is dominated by non-degenerate particles, and that
it balances $25\%$ of the gravitational energy $e_{\rm grav}$ at the circularization radius,
one finds a characteristic torus temperature
\begin{equation}
\label{eq:Tvir_def}
T_{\rm vir} \simeq 3\times 10^8 \left(\frac{e_{\rm int}}{4e_{\rm grav}}\right)\left(\frac{\mu}{1.75}\right)
                   \left( \frac{M_{\rm c}}{1.4M_\sun}\right)
                   \left( \frac{10^{9.3}\textrm{ cm}}{R_0}\right)\textrm{ K},
\end{equation}
where
\begin{equation}
\label{eq:mmw}
\mu = \left( Y_e + \sum_i \frac{X_i}{A_i}\right)^{-1}
\end{equation}
is the mean molecular weight, with $Y_e$ the electron fraction, and $\left\{X_i,A_i\right\}$ the mass 
fraction and atomic number of ionic species $i$, respectively.

The characteristic timescale for matter to accrete may be estimated by the
viscous time\footnote{Note that $t_{\rm visc}$ evaluated at large radii
in the disk generally underestimates the true accretion timescale onto the
central object since the net accretion rate usually decreases with decreasing
radii (e.g.,~$\dot{M} \propto r$) in most models of RIAFs.  Equations
(\ref{eq:tacc}) and (\ref{eq:mdot_out}) nevertheless represent useful
characteristic values.}
\begin{eqnarray}
\label{eq:tacc}
t_{\rm visc}& \simeq &\alpha^{-1}\left(\frac{R_{0}^{3}}{GM_{\rm c}}\right)^{1/2}\left(\frac{H_{0}}{R_{0}}\right)^{-2}\nonumber \\
&\sim& 2600{\rm\,s} \left(\frac{0.01}{\alpha}\right)\left(\frac{R_{0}}{10^{9.3}{\,\rm cm}}\right)^{3/2}
\left(\frac{1.4M_\sun}{M_{\rm c}}\right)^{1/2}
\left(\frac{H_0}{2R_{0}}\right)^{-2},\nonumber\\
\end{eqnarray}
where $\alpha$ parametrizes the disk viscosity ($\S\ref{s:viscosity}$),
resulting in a characteristic accretion rate
\begin{eqnarray}
\label{eq:mdot_out}
\dot{M}_{0}  &\sim& \frac{M_{\rm WD}}{t_{\rm visc}} \nonumber \\
&\sim & 2\times 10^{-4}M_{\sun}{\,\rm s^{-1}}\left(\frac{\alpha}{0.01}\right)
\left(\frac{M_{\rm WD}}{0.6M_{\sun}}\right)\left(\frac{R_{0}}{10^{9.3}{\,\rm cm}}\right)^{-3/2}\nonumber\\
& &\qquad\qquad\qquad\times \left(\frac{M_{\rm c}}{1.4M_\sun}\right)^{1/2} \left(\frac{H_0/R_{0}}{0.5}\right)^{2}.
\end{eqnarray}

For the thermodynamic conditions above, the opacity is dominated
by electron scattering. The timescale for photons to diffuse out of the disk midplane is then
\begin{equation}
t_{\rm diff} \simeq \frac{\kappa_{\rm es}\bar{\rho}}{c} H^2_{0} 
\sim 10^5 \left( \frac{M_{\rm WD}}{0.6M_\sun}\right) 
                    \left( \frac{10^{9.3}\textrm{ cm}}{R_0}\right)\left(\frac{H_0/R_0}{0.5}\right)^2\textrm{ yr},
\end{equation}
where $\kappa_{\rm es}$ is the electron scattering opacity.  Note that since
$t_{\rm diff}$ is much longer than the timescale for disk formation ($\sim
t_{\rm orb}$) or viscous evolution ($\sim t_{\rm visc}$), this implies that the
disk is highly radiatively inefficient and hence radiation losses can be
neglected.  

We limit our simulations to regions of the disk external to the radius $R_{\rm
in} \sim 10^{7}$ cm $\sim 0.01R_{0}$, a choice made primarily for computational
expediency.  This cutoff is justified because nuclear burning primarily occurs
exterior to this radius, with only {\it endothermic} dissociation happening
inside this point \citepalias{M12}.  

Neutrinos can in principle also cool the disk.  The
timescale for neutrino cooling near the disk circularization radius $R_{0}$ due
to $e^{-}/e^{+}$ pair annihilation is estimated to be $t_\nu \sim 5\times
10^{5}$ yr at the fiducial densities and temperatures given in equations
(\ref{eq:rho_ave}) and (\ref{eq:Tvir_def}), again much longer than the
timescale for disk evolution.  However, as matter flows inwards to smaller
radii where the temperature is higher, neutrino cooling is increasingly
important, possibly even causing the disk to transition to a geometrically thin
configuration (e.g.,~\citealt{DiMatteo+02}).  A straightforward calculation
shows that a thin disk obtains inside a critical radius $R_{\nu}$ given by
(\citealt{Chen&Beloborodov07}; see also \citealt{Metzger+08}, their eq.~[11]) 
\be
R_{\nu} \approx 2\times 10^{5}{\rm cm}\left(\frac{\dot{M}}{10^{-4}M_{\odot}\,{\rm s}^{-1}}\right)^{6/5}
\left(\frac{M_{\rm c}}{1.4 M_{\odot}}\right)^{-3/5}\left(\frac{\alpha}{0.01}\right)^{-2}.
\label{eq:Rnu}
\ee
This radius assumes cooling due to electron/positron capture on
free nuclei and hence is only strictly valid if the temperature is sufficiently
high to photodisintegrate heavy nuclei; however, it still serves as a useful
estimate of the radial scale at which neutrino cooling affects the global
thermodynamics of the disk.
If the characteristic accretion rate near the outer edge of the disk
(\ref{eq:mdot_out}) is similar to that at smaller radii in the disk (as may
{\it not} be valid for RIAFs), then from equation (\ref{eq:Rnu}) one infers
that a thin disk is only possible at radii $\ll 10^{7}$ cm.  Although neutrino cooling is
unlikely to significantly alter the 
disk thermodynamics in our computational domain,
we nevertheless include this effect in most
calculations ($\S\ref{s:time_dependent}$) to 
compensate for the exclusion of radiation pressure (\S\ref{s:nuclear_energy}).

\subsection{Angular Momentum Transport}
\label{s:viscosity}

Transport of angular momentum in a fully ionized disk is thought to be mediated
primarily via Maxwell stresses associated with the MRI.  If the initial
magnetic field of the WD is strong and the torus forms with a strong poloidal
field, then a magnetocentrifugal wind can also carry away angular momentum
(e.g., \citealt{blandford82}; \citealt{stone2001}).  The interior magnetic
fields of WDs are not well constrained observationally, but the measured
exterior fields range from $\lesssim$ 10 kG in most WDs, up to several hundred
MG in a small population of highly magnetized WDs
\citep{Wickramasinghe&Ferrario00}.  These fields are much less than the value
$\sim 10^{12}$ G that would reach equipartition with the gas pressure in the
torus,\footnote{The large-scale field may be amplified somewhat during 
disk formation via linear field winding, but this is only likely to enhance the
toroidal field.} suggesting that angular momentum loss to winds may be
neglected.

Assuming that the magnetic field is weak, the evolution of the torus should
resemble that found by previous three dimensional MHD simulations of RIAFs onto
black holes at large distances from the event horizon
(\citealt{hawley2001,igumenshchev2003}).  Unfortunately, MHD simulations are
computationally expensive.  One reason is that toroidal magnetic fields
amplified by the MRI tend to rise buoyantly into low density regions above the
disk midplane (e.g., \citealt{davis2010,shi2010}).  This causes the Alfv\'en
speed to be very large, thus limiting the dynamic range in radii or evolution
time that can be explored due to a small Courant-Friedrichs-Lewy (CFL) timestep
\citep{stone2001}.  This constraint becomes especially severe given the need to
resolve the most unstable mode of the MRI for a reliable saturation amplitude
(e.g., \citealt{hawley1995}).

A less costly alternative, which we adopt, is to perform hydrodynamic
simulations with a suitably chosen anomalous shear stress.  There are obvious
caveats to this approach.  First, the MRI converts orbital kinetic energy
directly into turbulent magnetic and kinetic energy, with the ensuing stresses
being responsible for the transport of angular momentum. In contrast, viscous
heating converts orbital energy into heat, which then drives convection through
entropy gradients \citep{hawley2001}.  This causes the hydrodynamic models to
achieve a state of marginal convective stability \citep{stone1999}, whereas in
MHD the very same stability criterion must be violated in order for the MRI to
operate \citep{hawley2001}.  The relative spatial distribution of specific
angular momentum and entropy thus differs between the hydrodynamic and MHD cases.  
In addition, the spatial and temporal distribution of heating is very different
between MRI-driven magnetohydrodynamic turbulence and shear viscosity
(e.g.,~\citealt{Hirose+06}).  This could potentially alter the thermodynamic
structure of the disk and hence the spatial distribution of nuclear burning.  

Despite these caveats, purely hydrodynamic simulations capture important
aspects of MHD simulations.  First, the time-averaged mass fluxes in the disk
midplane have the same qualitative form, with strong inflow and outflow nearly
canceling each other out, resulting in 
a net accretion rate that is approximately constant with 
radius \citep{stone1999,stone2001,hawley2001}.  Second, 
the radial scalings of time-averaged quantities
(e.g.,~density and temperature) found by MHD simulations can be matched by
hydrodynamic simulations if the functional form of the shear stress is suitably
chosen \citep{stone2001}.  These scalings do not differ substantially between
two- and three dimensions \citep{hawley2001}.

The experimental approach of this study demands the ability to simulate accretion
torii over a large dynamic range in spatial distances and timescales.  
We thus begin our study of NuDAFs by performing hydrodynamic
simulations with an anomalous viscous stress.  A similar approach was adopted
recently by \citet{Schwab+12} in studying the evolution of accretion disks
created by WD-WD mergers.  In order to best evaluate the uncertainties
introduced by this approach, we adopt several functional forms for the
kinematic viscosity, each of which lead to different transient and
quasi-steady-state outcomes.  

The first parameterization 
is that which makes the ratio of viscous to orbital times independent
of radius,
\begin{equation}
\label{eq:viscosity_constant}
\nu_{\rm ct} = \tilde\nu_0 \sqrt{GM_{\rm c}R_0}\left( \frac{r}{R_0}\right)^{1/2},
\end{equation} 
where $\tilde\nu_0$ is a dimensionless constant.  This form was found by \citet{stone1999} to yield
a self-similar behavior in the inner regions of the disk, with $\rho\propto
r^{-1/2}$ and $T\propto r^{-1}$.  The relationship between $\tilde\nu_0$ and the
ratio of timescales is
\begin{equation}
\label{eq:nu0_norm}
\tilde\nu_0 \simeq 1.6\times 10^{-3} \left[\frac{t_{\rm visc}/t_{\rm orb}}{100}\right].
\end{equation}
We employ this prescription to compare with analytic expectations for the
dynamical importance of burning (\S\ref{s:nuclear_energy}).

We also employ a
\citet{shakura1973} parameterization
\begin{eqnarray}
\label{eq:viscosity_alpha}
\nu_\alpha & = & \alpha \frac{c_s^2}{\Omega_K}\nonumber\\
             & = & \alpha \left( \frac{c_s^2}{GM_{\rm c}/r}\right) \sqrt{GM_{\rm c}R_0}\left( \frac{r}{R_0}\right)^{1/2}
\end{eqnarray} 
where $\Omega_K$ is the Keplerian frequency and $c_s^2 = \gamma p /\rho$ is the adiabatic
sound speed.
The functional form of $\nu_\alpha$ differs from $\nu_{\rm ct}$ in the additional dependence on the
ratio of thermal to gravitational energies. At the circularization radius, the numerical
value of $\tilde\nu_0$ in equation~(\ref{eq:nu0_norm}) corresponds to  $\alpha \sim 0.01$, with
exact values depending on the thermal content of the disk.

Finally, to connect with MHD calculations, we employ the prescription
of \citet{stone1999}
\begin{equation}
\label{eq:viscosity_spb}
\nu_{\rm spb} = \tilde\nu_0 \sqrt{GM_{\rm c}R_0}\,\left(\frac{\rho}{\rho_1}\right),
\end{equation}
where $\rho_1$ is a reference density, taken to be
the maximum value in the initial condition (\S\ref{s:initial_conditions}). 
This parameterization yields radial scalings in agreement with axisymmetric
MHD simulations with low initial poloidal fields \citep{stone2001}.

Following \citet{stone1999}, we include only the azimuthal components of the viscous stress 
tensor $\mathbb{T}$ in our simulations. 
\begin{eqnarray}
\label{eq:trphi_def}
T_{r\phi}      & = & \rho \nu\,\frac{r}{\sin\theta}\frac{\partial}{\partial r}\left(\frac{\ell_z}{r^2} \right)\\
T_{\theta\phi} & = & \rho \nu\,\frac{\sin\theta}{r^2}\frac{\partial}{\partial\theta}\left(\frac{\ell_z}{\sin^2\theta} \right).
\end{eqnarray}
Including all components of the stress would make the fluid truly viscous, and suppress 
convection in the poloidal direction (e.g., \citealt{igumenshchev1999}). 
Our simulations, like previous hydrodynamic studies, mimic 
turbulent angular momentum transport via thermally-driven convection. 
Suppressing this effect would suppress the very mechanism that gives plausibility to a
hydrodynamic modeling. 

Given the potentially large mass of the disrupted WD relative to that of the
central NS/BH, it is also possible that the disk will be susceptible to
gravitational instabilities, which could contribute to the transport of
angular momentum via spiral density waves (e.g.,~\citealt{Larson84}).  
We neglect this possibility here,
since pressure waves may stabilize such instabilities (the disk is quasi-virial
and geometrically thick). Also, our analysis does not include self-gravity.
However, to the extent that such waves are present and
transport angular momentum outwards, 
some of the qualitative effects of self-gravity may be captured by our 
prescription for angular momentum transport.

\subsection{Microphysics}
\label{s:nuclear_energy}

For a given initial composition of
the torus, we focus on the reaction activated first at large radii in the disk,
since energy input from this
reaction is the most important relative to the local rate of viscous heating
\citepalias{M12}. 

We focus mostly on the reaction
$^{12}$C($^{12}$C,$\gamma$)$^{24}$Mg, relevant to disks formed from mid-mass
C-O WDs.  We also calculate a few models that use the triple-alpha reaction, 
as would be appropriate for disks arising from low-mass He WDs or He stars.
Our analysis is also applicable to $\alpha-$capture 
reactions [e.g.,~$^{4}$He($^{16}$O,$\gamma$)$^{20}$Ne], which are relevant 
for a hybrid He-C-O WD disk or to the outer layers of a collapsing
Wolf-Rayet star (Woosley $\&$ Heger 2006).  
We exclude the parameter regime relevant to massive WDs with O-Ne composition, 
because the small circularization radii 
$R_{0}$ (eq.~[\ref{eq:r_circ}]) and high disk temperatures $T_{\rm vir}$
(eq.~[\ref{eq:Tvir_def}]) suggest
that in these systems burning begins during the circularization process itself \citepalias{M12}.

To better enable analytic comparisons, we adopt a generic power-law 
form for the 
$^{12}$C($^{12}$C,$\gamma$)$^{24}$Mg reaction rate
\begin{eqnarray}
\label{eq:Qdot_parametric}
\dot X_{\rm f} & = -A X_{\rm f}^2\rho\, T^\beta,
\end{eqnarray}
where $X_{\rm f}$ is the mass fraction of fuel (carbon),
$A$ is a normalization constant, and $\beta$ results
from expanding the analytic expression for the reaction rate as a Taylor series
in temperature (e.g.,~\citealt{Kippenhahn&Weigert94}) about the point at which
the burning time and dynamical time (or viscous time) are comparable in the
disk midplane.  
In the temperature range $(0.6-1.2)\times 10^9$~K, this procedure 
yields $\beta=29$.
In some of our models we employ the
full functional form of each reaction \citep{caughlan1988},
suitably converted to code units given the parametric equation of state.
Details of this procedure are given Appendix~\ref{s:units_reactions}. 

Associated with each reaction is the specific nuclear binding energy released,
\begin{equation}
\label{eq:enuc_definition}
\varepsilon_{\rm nuc} = Q\,X_{\rm f}/m_{a},
\end{equation} 
where $Q$ is the total energy released
per reaction and $m_{a}$ is the mass of the reaction product.  In several
simulations we artificially suppress
the value of $Q \propto \varepsilon_{\rm nuc}$
from its true physical value $Q_{0}$ in order to explore how the outcome of
NuDAF evolution depends the amount of nuclear energy released.

For simplicity, we use an ideal gas equation of state with a single adiabatic
index $\gamma = 5/3$.  As shown in \citetalias{M12}, gas pressure exceeds radiation pressure
close to the disk midplane during phases of evolution when most of the mass
accretes (i.e.~on timescales $t \gtrsim t_{\rm visc}$).  However, radiation
pressure contributes an increasingly larger fraction of the total pressure at
smaller radii in the disk. It dominates over gas pressure when the
density is low at early times prior to when the accretion rate achieves a
steady-state.  Neglecting radiation pressure results in an overestimate of the
temperature (and hence the rate of nuclear reactions), potentially producing
unphysical prompt detonations in some of our simulations, an issue we discuss
further in $\S\ref{s:detonations}$.  For completeness, we also include neutrino 
cooling as parameterized by
\citet{itoh1996}.\footnote{\url{http://cococubed.asu.edu/code\_pages/codes.shtml}}

\section{Numerical Method}
\label{s:numerical}

\subsection{Equations Solved}
\label{eq:}

Spherical polar coordinates in axisymmetry $(r,\theta)$ are adopted, with the origin at the center of the 
neutron star. We solve the equations of mass, momentum, energy, and
chemical species conservation, with source terms due to gravity, shear viscosity, and nuclear reactions: 
\begin{eqnarray}
\label{eq:mass_conservation}
\frac{\partial \rho}{\partial t} + \nabla \cdot (\rho\mathbf{v}_p) & = & 0\\ 
\label{eq:momentum_conservation}
\frac{\totd \mathbf{v}_p}{\totd t}  & = &
\mathbf{f}_{\rm c}-\frac{1}{\rho}\nabla p  -\frac{GM_{\rm c}}{r^2}\,\hat r \\
\label{eq:angular_conservation}
\rho\frac{\totd \ell_z}{\totd t} & = & r\sin\theta\,(\nabla\cdot\mathbb{T})_\phi\\
\label{eq:energy_conservation}
\rho\frac{\totd e_{\rm int}}{\totd t} + p\nabla\cdot\mathbf{v}_p 
& = & \frac{1}{\rho\nu}\mathbb{T}:\mathbb{T} + \rho\left(\dot{Q}_{\rm nuc} - \dot{Q}_{\rm cool}\right)\\
\label{eq:fuel_evolution}
\frac{\totd X_{\rm f}}{\totd t} & = & \dot{X}_{\rm f} = -\frac{\dot Q_{\rm nuc}}{Q/m_{\rm f}},
\end{eqnarray}
where $\rho$, $p$, $\mathbf{v}_p=v_r\hat r + v_\theta\hat\theta$, and $e_{\rm int}$ are the 
fluid density, pressure, poloidal velocity,
and internal energy, respectively, and $\totd/\totd t\equiv \partial/\partial t + \mathbf{v}_p\cdot\nabla$ is
the Lagrangian differential operator. 

The specific angular momentum along the symmetry axis satisfies $\ell_z = r\sin\theta v_\phi$,
with $v_\phi$ the azimuthal velocity. The corresponding centrifugal force in the 
poloidal direction is
\begin{equation}
\mathbf{f}_c = \frac{\ell_z^2}{r^3\sin^3\theta}\left[\sin\theta\hat r +\cos\theta\hat\theta\right].
\end{equation}

The system of equations~(\ref{eq:mass_conservation})-(\ref{eq:fuel_evolution}) is closed with an ideal gas
equation of state with adiabatic index $\gamma=5/3$, so that $p = (\gamma-1)\rho e_{\rm int}$.
The nuclear abundances are constrained by baryon number conservation, $X_{\rm f}+X_{\rm a} + X_{\rm x}=1$,
where $X_{\rm a}$ and $X_{\rm x}$ denote the mass fractions of ash and inert species, respectively.
The nuclear energy generation rate $\dot Q_{\rm nuc}$ is given by one of the 
full reaction rates described in Appendix~\ref{s:units_reactions}, or by a power-law
approximation using equations~(\ref{eq:Qdot_parametric}) and (\ref{eq:fuel_evolution}).
$\dot{Q}_{\rm cool}$ is the neutrino energy loss rate per unit mass.

Throughout this paper, we express quantities in terms of the Keplerian velocity at the
circularization radius $v_{\rm k0} = (GM_{\rm c}/R_0)^{1/2}$, and the orbital time at the same location,
$t_{\rm orb} = 2\pi\,R_0^{3/2}/(GM_{\rm c})^{1/2}$.

\subsection{Initial Conditions}
\label{s:initial_conditions}

The initial condition is a constant entropy and specific angular momentum torus
(e.g., \citealt{PP84}) with uniform chemical composition. A realistic
merger will produce a torus with a more general radial distribution of angular momentum,
since the disruption process is not instantaneous (e.g., \citealt{fryer1999}).  
We ignore this complication in the interest of clarity. The implications
of this assumption in light of our results are discussed in \S\ref{s:merger_discussion}.

Following \citet{stone1999}, the torus density is normalized to its maximum
value $\rho_{\rm max}$, thereby fixing the polytropic constant in terms of the
adiabatic index and the torus distortion parameter $d$. The resulting initial
density distribution is
\begin{equation}
\label{eq:density_distribution}
\left(\frac{\rho}{\rho_{\rm max}}\right)^{\gamma-1} 
= \frac{2d}{d-1}\left[\frac{R_0}{r}-\frac{1}{2}\left(\frac{R_0}{r\sin\theta}\right)^2 
			     			-\frac{1}{2d}\right].
\end{equation}

The distortion parameter $d$ is a measure of the internal energy content of the 
torus. The maximum ratio of internal to gravitational energy occurs at the point 
of maximum density ($r=R_0$, $z=0$),
\begin{equation}
\label{eq:internal_energy_grav}
\frac{e_{\rm int,max}}{GM/R_0} = \frac{1}{2\gamma}\frac{d-1}{d}.	
\end{equation}
Given the angular momentum distribution $j(r)$ of the torus, the value of $d$
is in principle fully determined by the properties of the disrupted binary and
energy conservation.  Unfortunately, the uncertainty in $j(r)$ and the
complication of self-gravity (which is not included in the initial torus solution) 
impede a precise determination of $d$.  In the absence
of more detailed information, we adopt $d=1.5$ as a fiducial value for most
of our models. For $\gamma=5/3$, this yields a maximum ratio of internal 
to gravitational energy (eq.~[\ref{eq:internal_energy_grav}]) of $10\%$. 
We also use $d=\{1.2,3\}$ in some models, to study the sensitivity
of our results to this choice. The integrated mass distributions so obtained bracket
the results of \citet{fryer1999} for WD-BH disks.

A physical temperature is obtained by assuming that the pressure is provided by
a non-relativistic ideal gas, and using a mean molecular weight (eq.~[\ref{eq:mmw}])
appropriate to the initial WD composition. Similarly, a physical density
is obtained by multiplying the average disk density (eq.~[\ref{eq:rho_ave}])
by the ratio of the maximum to average density in the initial torus. This ratio
is a function of $d$ and $\gamma$ only. Details of this unit conversion are 
provided in Appendix~\ref{s:units_reactions}. 

The torus is surrounded by a low-density isothermal atmosphere,
\begin{equation}
\label{eq:isothermal_atmosphere}
\rho_{\rm at}(r) = \rho_{\rm at,0}\,
\exp\,\left\{\frac{GM}{c^2_{\rm at}\, r_{\rm in}}\left(\frac{r_{\rm in}}{r}-1 \right) \right\},
\end{equation}
where $r_{\rm in}$ is the inner radial boundary of the computational domain,
and $c^2_{\rm at} = p_{\rm at,0}/\rho_{\rm at,0}$ is the isothermal sound speed 
of the atmosphere. This surrounding medium has the advantage of being 
stably-stratified. We find that a constant density atmosphere, such as 
that used by \citet{stone1999}, generates excessive numerical noise near 
the inner boundary. The normalization
constant is chosen to be much smaller than the bulk of the torus density,
and an order of magnitude below the density at which we cut off source
terms (\ref{s:time_dependent}).
At radii $r\gg r_{\rm in}$, the density asymptotes to 
$\rho_\infty = \rho_{\rm at}\exp{[-GM_{\rm c}/(c^2_{\rm at} r_{\rm in})]}$.
In most models, we adopt $\rho_{\rm at,0}/\rho_{\rm max} = 10^{-6}$
and $c^2_{\rm at} = 25GM_c/R_0$. The temperature is chosen so that the resulting
density scale height at the inner boundary is resolved with at least 5 cells
in radius.
This choice of atmospheric parameters causes a negligible effect on the
torus evolution even though the atmosphere itself is unbound. In a few
models, we increase the background density to $\rho_{\rm at,0}/\rho_{\rm max} = 10^{-4}$
to obtain smoother evolution of the inner edge of the torus during the initial
infall. This higher value artificially slows down ejected material, however, and is kept
for comparison purposes only.

\subsection{Time-dependent evolution}
\label{s:time_dependent}

We use FLASH3.2 \citep{dubey2009} to evolve the system of 
equations~(\ref{eq:mass_conservation})-(\ref{eq:fuel_evolution}) with the
dimensionally-split version of the Piecewise Parabolic Method (PPM, \citealt{colella84}).
The public version of the code has been modified to allow for a non-uniformly 
spaced grid in spherical polar coordinates as described in \citet{F12}.

The specific angular momentum is included as an advected scalar 
subject to viscous source terms (eq.~[\ref{eq:angular_conservation}]). To take
advantage of the finite-volume character of the hydrodynamic method, the diffusion
operator is recast as
\begin{equation}
\label{eq:flux_divergence}
r\sin\theta\, (\nabla\cdot \mathbb{T})_\phi = \nabla\cdot\mathbf{F}_\ell,
\end{equation}
where
\begin{equation}
\label{eq:angz_flux}
\mathbf{F}_\ell = r\sin\theta\left(T_{r\phi}\hat r + T_{\theta\phi}\hat \theta \right)
\end{equation}
is the angular momentum flux vector. This diffusive flux is added to the advective flux obtained
from the Riemann problem solution at each cell interface. The combined flux is then used
to update the volume-averaged value of $\ell_z$ (see \citealt{lindner2010} for a similar
implementation in cylindrical coordinates).
The internal energy is updated by simple operator-split addition of the scalar viscous 
energy generation rate (equation~{\ref{eq:energy_conservation}}).
Numerical stability requires the evolution 
time step to be lower than  
\begin{equation}
\Delta t_{\rm visc} = \frac{1}{2}\min\left(\frac{\Delta r^2}{\nu}, \frac{r^2\Delta\theta^2}{\nu}\right),
\end{equation}
where the minimization is carried out over the entire computational domain. This constraint on
the time step is imposed in addition to the CFL condition required by the hydrodynamic method.
The centrifugal force is included as part of the default treatment of `fictitious' forces 
that arise in curvilinear coordinates.

By default, FLASH3.2 evolves the total specific energy (internal plus kinetic). 
To avoid adding the additional source terms needed to account for the Eulerian rate of change
of the rotational kinetic energy, we only evolve internal energy
(equation \ref{eq:energy_conservation}).

The nuclear energy generation rate and neutrino cooling are 
also included via simple operator-split update of the internal energy.
We suppress nuclear burning inside shocks, and employ 
a standard treatment of multi-species advection \citep{plewa1999}.
To avoid artificial thermonuclear runaways, we impose an additional 
constraint on the time-step from nuclear burning
\begin{equation}
\Delta t_{\rm nuc} = 0.8 \frac{e_{\rm int}}{\dot Q_{\rm nuc}}.
\end{equation}
In practice, the simulation time-step is dominated by the CFL condition given
the atmospheric temperature and the grid resolution, being usually $10$ to $100$ times
smaller than $\Delta t_{\rm nuc}$.

\begin{deluxetable*}{lcccccccccccccc}
\tablecaption{Models Evolved\label{t:models}}
\tablewidth{0pt}
\tablehead{
\colhead{Model}&
\colhead{$M_{\rm WD}$} &
\colhead{$X_{\rm He}/X_{\rm C}/X_{\rm O}$\tablenotemark{a}} &
\colhead{$d$\tablenotemark{b}}          & 
\colhead{Reaction\tablenotemark{c}} &
\colhead{$Q/Q_0$\tablenotemark{d}} &
\colhead{Visc.\tablenotemark{e}} &
\colhead{$\tilde\nu_0$, $\alpha$\tablenotemark{f}} &
\colhead{$N_r$\tablenotemark{g}} &
\colhead{$\theta_{\rm p}$\tablenotemark{h}} &
\colhead{$r_{\rm in}/R_0$} &
\colhead{$R_0$} &
\colhead{$\rho_\infty$\tablenotemark{i}} &
\colhead{$\rho_0\tablenotemark{j}$} &
\colhead{$\nu$ Cool?} \\
\colhead{ } & \colhead{($M_\sun$)} & \multicolumn{8}{c}{ } & \colhead{($M_\sun$)} & 
\colhead{(cm)} & \colhead{($\rho_{\rm max}$)} & \colhead{($\rho_{\rm max}$)} & \colhead{ }
}
\startdata
COq000 & 0.6 & 0/0.5/0.5  & 1.5 & power-law   & 0.00    & ct & 0.0016  & 64 & 4 & $10^{-2}$ & $10^{9.3}$ & $10^{-5.7}$ & $10^{-3}$ & Yes\\
COq010 &     &            &     &             & 0.10    &    &         &    &   &           &            &             &           & \\
COq025 &     &            &     &             & 0.25    &    &         &    &   &           &            &             &           & \\
COq050 &     &            &     &             & 0.50    &    &         &    &   &           &            &             &           & \\
COq075 &     &            &     &             & 0.75    &    &         &    &   &           &            &             &           & \\
COq100 &     &            &     &             & 1.00    &    &         &    &   &           &            &             &           & \\
\noalign{\smallskip}
COq000\_nc & 0.6 & 0/0.5/0.5  & 1.5 & power-law   & 0.00    & ct & 0.0016  & 64 & 4 & $10^{-2}$ & $10^{9.3}$ & $10^{-7.7}$ & $10^{-3}$ & No\\
COq010\_nc &     &            &     &             & 0.10    &    &         &    &   &           &            &             &           &  \\
COq025\_nc &     &            &     &             & 0.25    &    &         &    &   &           &            &             &           &   \\
COq050\_nc &     &            &     &             & 0.50    &    &         &    &   &           &            &             &           &   \\
COq100\_nc &     &            &     &             & 1.00    &    &         &    &   &           &            &             &           &   \\
\noalign{\smallskip}
COq050\_HR   & 0.6 & 0/0.5/0.5 & 1.5 & power-law & 0.50 & ct  & 0.0016  & 128 & 2 & $10^{-2}$ & $10^{9.3}$ & $10^{-5.7}$ & $10^{-3}$ & Yes\\
COq050\_at      &     &            &     &             &      &     &         & 64  & 4 &     &            & $10^{-7.7}$ &           & \\
COq050\_$\rho_0$&     &            &     &             &      &     &         &     &   &     &            &             & $10^{-4}$ & \\
COq050\_dL      &     &            & 1.2 &             &      &     &         &     &   &     &            &             & $10^{-3}$ & \\
COq050\_dH      &     &            & 3.0 &             &      &     &         &     & 6 &     &            &             &           & \\
COq050\_$\nu$H  &     &            & 1.5 &             &      &     & 0.016   &     &   &     &            &             &           & \\
COq050\_spb     &     &            &     &             &      & spb & 0.008   &     & 4 &     &            &             &           & \\
COq050\_$\alpha$&     &            &     &             &   &$\alpha$& 0.0144  &     &   &     &            &             &           & \\
\noalign{\smallskip}
CO\_f1 & 0.6 & 0/0.5/0.5  & 1.2 & $^{12}$C+$^{12}$C   & 1.00 & ct & 0.016   & 64  & 4 & $10^{-2}$ & $10^{9.3}$ & $10^{-7.7}$ & $10^{-3}$ & Yes\\
CO\_f2 &     &            & 1.5 &              &      &    & 0.0016  &     &   &           &            &             &           &    \\
HE\_00 & 0.3 & 1/0/0 & 1.5 & 3$\alpha$    & 0.00 & ct & 0.016 & 64  & 4 & $10^{-2.5}$ & $10^{9.6}$ & $10^{-11.8}$ & $10^{-3}$ & Yes\\
HE\_f1 &     &       &     &              & 1.00 &    &       &     &   &             &            &              &           &   \\
\enddata
\tablenotetext{a}{Initial mass fractions of helium, carbon, and oxygen, respectively.}
\tablenotetext{b}{Distortion parameter controlling the initial thermal content of the torus (see eq.~[\ref{eq:internal_energy_grav}]).}
\tablenotetext{c}{Nuclear reaction rate employed (Appendix~\ref{s:units_reactions}).}
\tablenotetext{d}{Ratio of the nuclear energy released per reaction $Q$ to the true physical value $Q_{0}$.}
\tablenotetext{e}{Functional form of the kinematic viscosity
                  (eqs.~[\ref{eq:viscosity_constant}]-[\ref{eq:viscosity_spb}]).}
\tablenotetext{f}{Magnitude of viscosity $\tilde\nu_0$ or $\alpha$, as appropriate. 
		  The coefficient $\alpha$ is related to $\tilde\nu_0$ through equation~(\ref{eq:internal_energy_grav}). }
\tablenotetext{g}{Number of cells per decade in radius.} 
\tablenotetext{h}{Angular resolution at the polar axis} 
\tablenotetext{i}{Asymptotic atmospheric density (eq.~[\ref{eq:isothermal_atmosphere}]).}
\tablenotetext{j}{Cutoff density for source terms (eq.~[\ref{eq:density_cutoff}]).}
\end{deluxetable*}

The computational grid is logarithmically spaced in the radial direction, covering
four orders of magnitude in length. In most cases, it extends from 
$r_{\rm min} = 0.01R_0$ to $r_{\rm max}=100R_0$, although a few models use a smaller inner
radius (Table~\ref{t:models}). In the polar direction the grid covers the full range
of latitudes $([0^\circ,180^\circ])$, 
with cells having a constant ratio of sizes (e.g., \citealt{stone92}). We 
make this spacing symmetric relative to the equator, with the coarsest cells
next to the polar axis. Our standard resolution is $N_r = 64$ cells per decade
in radius. The angular spacing is chosen so that  
$\Delta\theta \simeq \Delta r /r \simeq 2^\circ$ at the equator, and 
$\Delta\theta_p = 4^\circ$ or $6^\circ$ next to the polar axis\footnote{We find that the funnel that develops
around the rotation axis can cause numerical problems if the angular 
resolution is too fine.}. We evolve one model at higher resolution, with $N_r = 128$ cells per
decade in radius, $\Delta\theta \simeq \Delta r /r \simeq 1^\circ$ at the equator,
and $\Delta\theta_p = 2^\circ$.
 
The boundary conditions at the polar axis are reflecting in all variables.
At the inner radial boundary, we impose a zero-gradient boundary condition
for the mass flux, subtracting the contribution from the isothermal atmosphere.
For a ghost cell with position $(r,\theta)$, the density, pressure, and radial velocity are set to
\begin{eqnarray}
\label{eq:boundary_conditions}
\rho(r,\theta) & = & \rho(r_1,\theta) - \rho_{\rm at}(r_1) + \rho_{\rm at}(r) \\
p(r,\theta)    & = & p(r_1,\theta) -p_{\rm at}(r_1) + p_{\rm at}(r)  \\
v_r(r,\theta)    & = & \left(r_1/r\right)^2\, v_r(r_1,\theta) 
\end{eqnarray}
if $v_r(r_1,\theta)<0$ and $\rho(r_1,\theta) > \rho_{\rm at}(r_1)$, where $r_1$ is the
radial coordinate of the center of the first active cell outside the
inner radial boundary. Otherwise the radial velocity in the ghost cells
is set to zero.
Because fluid elements associated with the disk normally
have densities much larger than that of the atmosphere, this is in effect
a standard outflow boundary condition. The additional terms ensure that
the isothermal atmosphere remains quiescent when no material is accreted.
The ghost cells for the internal energy are set in consistency to those
of the density and pressure. All other variables 
are given a zero radial gradient in
the ghost cells. This prescription is repeated at the outer radial boundary,
with the only modification being a reversal of the required sign of the
radial velocity for outflow.

To prevent excessive heating in regions of low density, 
we multiply all the energy and viscous source terms by a cutoff function $f(\rho)$.
Otherwise the CFL timestep can become prohibitively short in regions that are unimportant
for the disk evolution. The functional form of the cutoff is
\begin{equation}
\label{eq:density_cutoff}
f(\rho) =\left\{
\begin{array}{cc}
1                  & \rho > \rho_0\\ 
e^{(-\rho_0/\rho)} & \rho \le \rho_0,
\end{array}\right.
\end{equation}
where $\rho_0$ is a fiducial density, which we choose to be $\rho_0/\rho_{\rm
max} = 10^{-3}$.  In the initial torus configuration, $99.99\%$ of the mass is
included inside this contour.  We have verified that the evolution of
the relevant dynamic quantities is not very sensitive to the value of this cutoff
below a certain value (\S\ref{s:models}).  Although a low-density suppression of burning may seem
ad hoc, it can be motivated physically.  First, at low densities our assumed
ideal gas EOS overestimates the true temperature (and hence burning rate), since
we have neglected the effects of radiation pressure, which dominates the total
pressure in low density regions.  Also, in low density regions the photon
optical depth may become sufficiently small that radiative cooling (also
neglected in our simulation) may become relevant on the timescales of interest.

Numerical tests of the angular momentum evolution are presented 
in Appendix~\ref{s:code_tests}.

\subsection{Models Evolved}
\label{s:models}

\begin{figure*}
\includegraphics*[width=\textwidth]{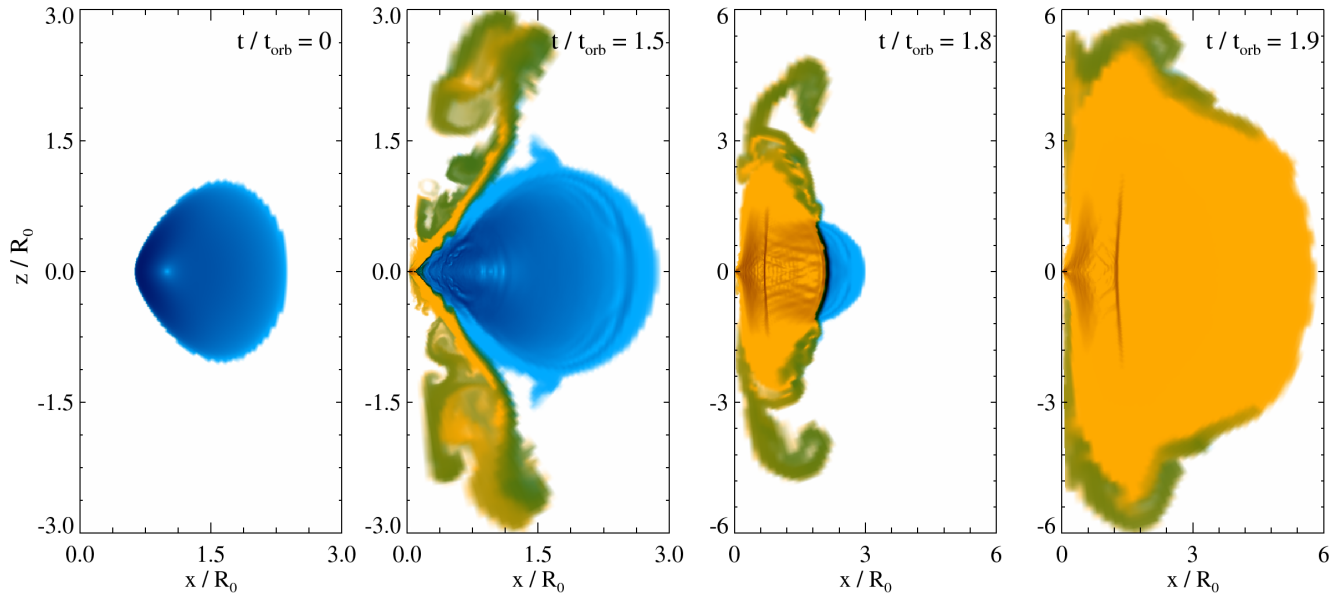}
\caption{Snapshots of model COq050\_HR, illustrating the main evolutionary
stages of the accretion of a tidally disrupted 0.6$M_{\odot}$ C-O WD onto a 1.4$M_{\odot}$ NS 
in a case that leads to a detonation.  The mass fractions of fuel (carbon) and ash
(magnesium) are shown in blue and yellow, respectively, with green regions indicating
mixed material. The shading is proportional to the density gradient.  From left to 
right, panels correspond to
(a) the initial condition; (b) establishment of a steady burning front at
$r_{\rm nuc}\sim 0.1R_0$ (eq.~[\ref{eq:rnuc_definition}]); (c) initiation of
detonation; and (d) shock expansion and partial unbinding of the torus.  Time
is shown in units of the orbital time at $r=R_0$.  Note that the spatial scale
is enlarged in the final two panels. An animated version of this figure is available
in the online version of the article.} 
\label{f:nudaf_truecolor}
\end{figure*}

We evolve a suite of models that systematically sample the parameter space of
disk properties relevant to 
WD-NS mergers. The complete set is shown in Table~\ref{t:models}.

Models are grouped according to the parameter being varied.
Most cases focus on a C-O WD of mass $M_{\rm WD} =
0.6M_\sun$ accreting onto a NS of mass $M_{\rm c} = 1.4M_{\odot}$. The
composition is $50\%$ carbon and $50\%$ oxygen.
The first two subsets of models focus on the effect of varying the 
energy released per nuclear reaction $Q$, expressed as a fraction
of its true physical value $Q_{0}$, using the power-law approximation
to the $^{12}$C$(^{12}$C,$\gamma$)$^{24}$Mg reaction (eq.~[\ref{eq:Qdot_parametric}]). 
One sequence includes neutrino cooling and the other does not (models ending in \emph{\_nc}). 

As we discuss below, most of our models with $Q \geq 0.5Q_{0}$ undergo large-scale
detonations, while those with $Q < 0.5Q_{0}$ instead undergo quiescent burning.
Since one of our goals is to understand what conditions are necessary for
detonation, we adopt the marginal case of the sequence with cooling (COq050)
for further parameter variation.
The third subset of models thus explores the effect of higher resolution (COq050\_HR),
torus distortion parameter (COq050\_dL and COq050\_dH), strength (COq050\_$\nu$H) and
functional form (COq050\_spb and COq050\_$\alpha$) of the shear stress, magnitude
of the atmospheric density (COq050\_at), and value of the cutoff density $\rho_0$ (COq050\_$\rho0$)
on the marginal model.

The fourth subset of models employs the full reaction rates to assess the likelihood
of detonation in disks formed from C-O  and He WDs. Model CO\_f1  employs the same parameters as our fiducial
C-O WD marginal model, but with the full strength of the energy release ($Q=Q_0$). Model
CO\_f2 decreases $d$ and increases $\tilde{\nu}_0$ to increase the chance of
explosion.  
Models He\_000 and He\_f1 explore the evolution of pure He
disks ($X_{\rm He} = 1$) under the influence of the triple-alpha reaction.
Parameters correspond to a $M_{\rm WD} = 0.3M_\sun$ WD accreting onto 
a NS of mass $M_{\rm c} = 1.4M_\sun$. The viscosity is chosen large, to 
maximize the likelihood of a detonation. The inner boundary of the domain
is chosen smaller than in the C-O case, because the ratio of burning to circularization radii
is expected to be smaller as well \citepalias{M12}.

\section{Results}
\label{s:results}

We begin the presentation of our results with a basic overview
($\S\ref{s:overview}$) of the impact of nuclear burning on the evolution of the
accretion disk, focusing on a model which detonates.  Then, in separate sections, we 
elaborate on exploding ($\S\ref{s:detonations}$) and quiescent cases
($\S\ref{s:quiescent}$).  The results of our simulations
are summarized in Table \ref{t:results}.

\vspace{0.3in}

\subsection{Overview}
\label{s:overview}

In the absence of nuclear burning, the disk evolves in a way similar to that in
the axisymmetric simulations of RIAFs by \citet{stone1999}.  The
initial torus develops a radial velocity through the action of viscous
stresses.  Material near the inner edge of the disk reaches the inner boundary
within the first few orbital periods as measured at $r=R_0$.  As more of the
disk spreads inward, viscous heating drives convective turbulence, resulting in
inward and outward mass fluxes of comparable magnitude.  Material is ejected in
a wide funnel around the rotation axis, but most of the ejecta has negative
energy and eventually falls back to the disk.  The accretion rate peaks on a
timescale which is comparable to a viscous time at $r=R_0$, before gradually
decreasing with time thereafter.  
The properties of torii without burning are discussed in relation to quiescent
NuDAFs in \S\ref{s:quiescent}.

The dynamical importance of nuclear energy generation at a given radius can be
quantified by the ratio of the specific nuclear energy released per reaction
$\varepsilon_{\rm nuc}$ (eq.~[\ref{eq:enuc_definition}]) to the local gravitational binding energy.
Given the steep temperature dependence of most reactions, fuel depletion and
energy deposition are generally concentrated in a narrow surface 
that crosses the midplane at a
characteristic radius $r_{\rm nuc}$ from the central object.  
If burning occurs subsonically, the magnitude of the density jump across
the burning region is given by the ratio of $\varepsilon_{\rm nuc}$ to
the specific enthalpy of the fluid $w$ (\citealt{landau}; see also Appendix~\ref{s:zeldovich}).
The latter is related to the specific gravitational binding energy by a factor
$\propto (H/r)^2$ \citepalias[e.g., ][]{M12}. 
Hence, one can use the ratio of $\varepsilon_{\rm nuc}$ to $w$ evaluated at $r = r_{\rm nuc}$, 
\begin{eqnarray}
\label{eq:psi_def}
\Psi & = & \frac{\varepsilon_{\rm nuc}}{w(r_{\rm nuc})}\\
\label{eq:psi_pp}
     & = & \frac{2d}{d-1}\frac{\varepsilon_{\rm nuc}}{GM_{\rm c}/R_0}\left(\frac{r_{\rm nuc}}{R_0}\right)^m
\end{eqnarray}
to parameterize the importance of nuclear reactions on the disk evolution.
Note that equation~(\ref{eq:psi_pp}) is specific to our initial conditions (\S\ref{s:initial_conditions});
it assumes that the temperature has a power-law dependence with radius, $T\propto r^{-m}$,
and makes use of equation~(\ref{eq:internal_energy_grav}) for normalization.
The values of $\Psi$ quoted in our results below are calculated using an
angle-averaged value of $w(r_{\rm nuc})$ obtained directly from our
simulations. Equation (\ref{eq:psi_pp}) is a reference 
value only. 

The location of $r_{\rm nuc}$ can be estimated by equating the 
dynamical time $r / v_{\rm k0}$ to the nuclear burning time 
$e_{\rm int}/\dot{Q}_{\rm nuc}$.
If the timescale for viscous heating is long compared to that for
vertical hydrodrostatic balance (the sound crossing time), then the disk
temperature $T$ approaches its virialized value $T_{\rm vir} \propto r^{-1}$
(eq.~[\ref{eq:Tvir_def}]). Otherwise it has a more general dependence $T\propto r^{-m}$. 
If the density scales as a power-law with radius
$\rho\propto r^n$, then using the power-law approximation to the
$^{12}$C($^{12}$C,$\gamma$)$^{24}$Mg reaction (eq.~[\ref{eq:Qdot_parametric}]),
and the power-law form for the viscosity prescription
(eq.~[\ref{eq:viscosity_constant}]), one finds that
\begin{equation}
\label{eq:rnuc_definition}
\frac{r_{\rm nuc}}{R_0} \simeq \left[\left(\frac{\varepsilon_{\rm nuc}}{e_{\rm int}}\right)\left(\frac{R_0}{v_{\rm k0}}\right)
                                     AX_{\rm f}\,\rho_0\,T_0^\beta \right]^{1/(m[\beta-1]+ n -3/2)}
\end{equation}
where the zero subscripts on thermodynamic quantities denote their values at
$r=R_0$. 
For C-O torii with $M_{\rm WD} = 0.6M_\sun$, $M_{\rm c}=1.4M_\sun$,
and $d=1.5$, we find $r_{\rm nuc}/R_0\simeq 0.1$, for $m=1$, and $n=2$. 
Equation~(\ref{eq:rnuc_definition}) is sensitive to the radial dependences
of the temperature and density, as well as to the temperature normalization at
the circularization radius.
The position of $r_{\rm nuc}$ will thus change in time as the disk approaches
a quasi steady-state configuration.

Assuming a carbon mass fraction $X_{\rm C}=0.5$ in $\varepsilon_{\rm nuc}$, 
a value of $r_{\rm nuc}\simeq 0.1$ 
yields $\Psi \simeq 1.8(Q/Q_0)$ for the power-law burning rate and a
central mass $M_{\rm c} = 1.4M_\sun$. 
Using the full functional form for the reaction rate requires solving for
$r_{\rm nuc}/R_0$ numerically. Table~\ref{table:psivalues} shows the result of
such a calculation for $^{12}$C($^{12}$C,$\gamma$)$^{24}$Mg and other
relevant reactions (e.g., $\alpha$-captures).  Note that $\Psi$ is similar in
all cases except for oxygen burning, which is lower by a factor $\sim 2$.

\begin{deluxetable}{lcccc}
\tablecaption{Dynamical Importance of Nuclear Reactions in RIAFs.\tablenotemark{a}\label{table:psivalues}}
\tablewidth{\columnwidth}
\tablehead{
\colhead{Reaction} &
\colhead{$Q_{0}$} &
\colhead{$X_{\rm He}/X_{\rm C}/X_{\rm O}$\tablenotemark{b}} &
\colhead{$r_{\rm nuc}$\tablenotemark{c}} &
\colhead{$\Psi$\tablenotemark{d}}
\\
 & \colhead{(MeV)} & & \colhead{($10^{8}$ cm)} & 
}
\startdata
$^{12}$C($^{12}$C,$\gamma$)$^{24}$Mg & 13.93 & 0.0/0.5/0.5 & 1.9 & 1.68 \\
$^{12}$C($\alpha$,$\gamma$)$^{16}$O & 7.16   & 0.2/0.5/0.3 & 2.5 & 1.74 \\
$^{16}$O($\alpha$,$\gamma$)$^{20}$Ne & 4.73  & 0.2/0.3/0.5 & 5.1 & 1.86 \\
$^{16}$O($^{16}$O,$\gamma$)$^{32}$S & 16.54  & 0.0/0.5/0.5 & 1.0 & 0.78 \\
\enddata
\tablenotetext{a}{Assuming a geometrically thick,
                  virialized torus with $R_{0} = 10^{9.3}$ cm; $\rho \propto r^{-1/2}$ 
                  ($\dot{M} \propto r$); viscosity parameter $\alpha = 0.01$; $H/r = 0.3$, and
                  $M_{\rm c} = 1.4M_{\odot}$.}
\tablenotetext{b}{Initial mass fractions}
\tablenotetext{c}{Burning radius at which half the initial fuel is consumed.  Its value is 
calculated numerically using the full functional form of the reaction rate.}
\tablenotetext{d}{Ratio of $\varepsilon_{\rm nuc}$ to the enthalpy $w = (\gamma-1)^{-1}(H/r)^2\,GM_{\rm c}/r_{\rm nuc}$
		  of the flow (eq.~[\ref{eq:psi_def}]).}
\end{deluxetable}

\begin{figure}
\centering
\includegraphics*[width=\columnwidth]{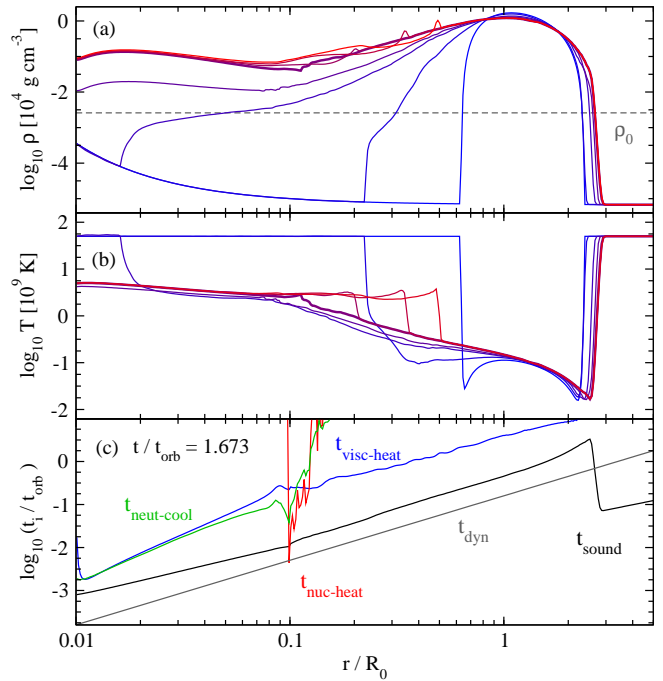}
\caption{{\it Panels (a) and (b):}  Angle-averaged density $\rho$ and temperature
$T$ as a function of radius for the model COq050\_HR
(Fig.~\ref{f:nudaf_truecolor}), respectively.  Averages are density-weighted and taken within
$45$ degrees of the equator.  From blue to red, solid curves correspond to
times $t/t_{\rm orb} = \{0,0.5,1,1.2,1.673,1.68, 1.69,1.7\}$.  The horizontal
dashed gray line shows the cutoff density $\rho_0$ below which nuclear burning
is artifically suppressed (eq.~[\ref{eq:density_cutoff}]). {\it
Panel (c):}  Various timescales in the disk as a function of radius at the time
$t/t_{\rm orb} = 1.673$ just prior to detonation (compare with
Fig.~\ref{f:nudaf_ignition}).  Timescales $t_i$ shown include: dynamical $t_{\rm dyn}
= r/v_k(r)$ (grey); sound crossing $t_{\rm sound} = r/c_s$ (black);
viscous heating $t_{\rm visc-heat}=e_{\rm int}/\dot{Q}_{\rm visc}$ (blue); 
nuclear burning $t_{\rm burn-heat}=e_{\rm int}/\dot{Q}_{\rm nuc}$ (red); 
and neutrino cooling $t_{\rm neut-heat}=e_{\rm int}/\dot{Q}_{\rm cool}$
(green).}
\label{f:rhoT_timescales}
\end{figure}

The qualitatively new behavior introduced by nuclear burning is illustrated
in Figure~\ref{f:nudaf_truecolor}, which
shows several snapshots in the evolution of the
C-O torus in our fiducial high resolution model COq050\_HR with $Q = 0.5Q_{0}$
($\Psi \simeq 0.6$ measured at the moment of final detonation, see Table~\ref{t:results}). 
Initially, the disk is composed entirely of fuel (carbon).
During the first orbital period of evolution, the inner edge of the disk moves
inwards and increases in temperature, before igniting at some radius initially
smaller than the fiducial burning radius $r_{\rm nuc}$.  A narrow burning front
is established which moves outwards and then settles at $r \approx r_{\rm nuc}$
on a (local) dynamical time.  
Turbulence is generated outside of the burning front,
triggering a thermonuclear runaway and detonation at the time $t =
1.67t_{\rm orb}$. 

After a time $t = 1.9t_{\rm orb}$, the detonation has successfully propagated
out through the entire torus.  At this point approximately $99\%$ of the
initial WD mass ($\sim 0.6M_{\odot}$) has become unbound.  The net energy in
this material is $E_{\rm ej} \simeq 3\times 10^{50}$ ergs, 
and has a
mass-weighted radial velocity of $\bar{v}_{\rm ej} \simeq 6000$ km s$^{-1}$.  A
fraction 5\% of this mass becomes bound and remains as a remnant disk.  In
$\S\ref{s:discussion}$ we discuss the possible observational signatures of this
supernova-like explosion.     

\begin{figure*}
\includegraphics*[width=\textwidth]{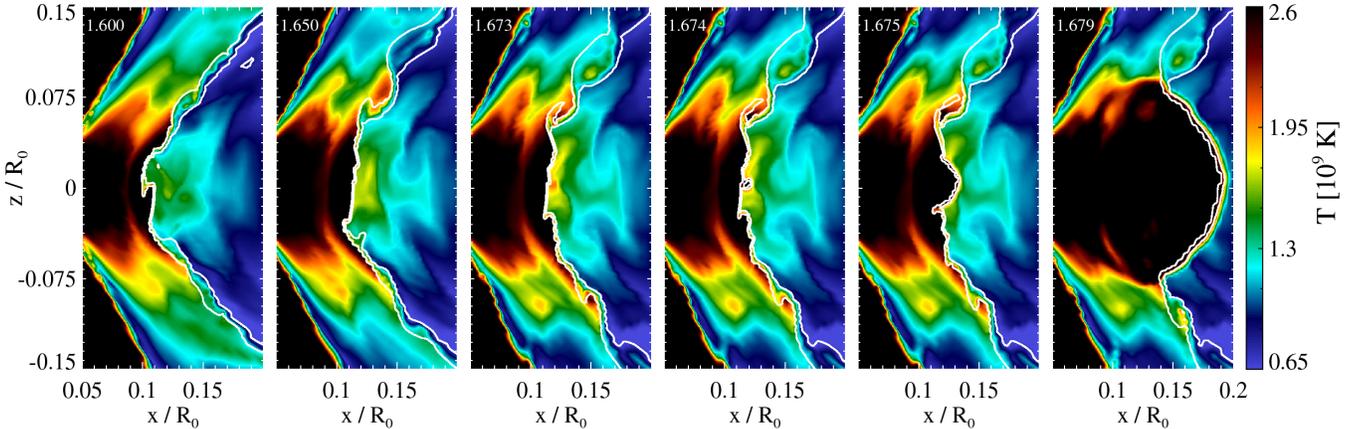}
\caption{The effect of turbulence near the burning front on the onset of
detonation in the model COq050\_HR.  Each panel shows snapshots of the
temperature, with the time in units of the orbital time at $r=R_0$ displayed in
the upper left corner.  White contours correspond to
fuel mass fractions $X_{\rm f} = \{0.1,0.4\}$.
From left to right, the panels show:
quasi-steady-state burning front ($t/t_{\rm orb} = 1.600$), formation of an
eddy by turbulence ($t/t_{\rm orb} = 1.650$), hot spot prior to burning
($t/t_{\rm orb} = 1.673$), hot spot after burning ($t/t_{\rm orb} = 1.764$),
and propagation of detonation ($t/t_{\rm orb} = 1.675$ and 1.679).  An animated
version of this figure is available in the online version of the article.}
\label{f:nudaf_ignition}
\end{figure*}

\begin{figure}
\includegraphics*[width=\columnwidth]{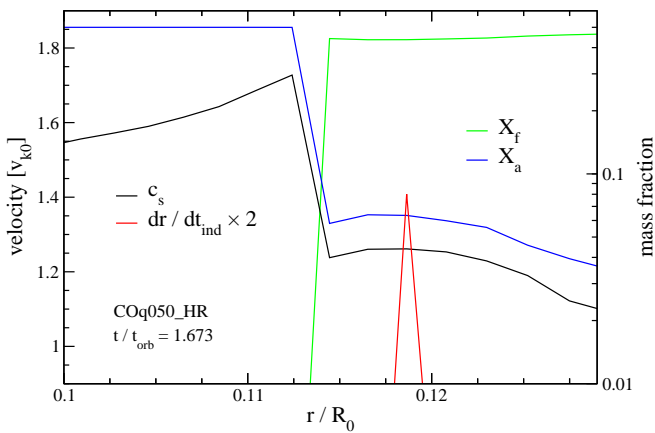}
\caption{Radial profile of quantities angle-averaged over the hot spot that triggers a detonation at time
$t=1.673t_{\rm orb}$ in the model COq050\_HR (Figure~\ref{f:nudaf_ignition}).
Shown are the mass fraction of fuel (green), ash (blue), the sound
speed (black), and the inverse of the induction time gradient (red). Mixing
of ash upstream of the front is responsible for the increased temperature.
This heating results in the condition $\totd r/\totd t_{\rm ind} \sim c_{s}$ for
spontaneous detonation (Appendix \ref{s:zeldovich}) being approximately
satisfied across the hot spot. The induction time is defined here
as $t_{\rm ind} = e_{\rm int}/[\beta\,\dot{Q}_{\rm nuc}]$, where $\beta=29$ is
the temperature exponent of the reaction rate (eq.~[\ref{eq:Qdot_parametric}]).}
\label{f:spot_profiles}
\end{figure}

Our calculations show that detonations can occur in NuDAFs ranging from
\emph{microexplosions} (e.g., \citealt{lisewski2000}) which remain localized,
to complete consumption of the initial disk.  A large-scale detonation capable of unbinding the
disk, and occurring within the first few orbital periods of evolution, appears
to be a robust outcome if $\Psi$ is above a critical value $\Psi_{\rm crit}
\sim 1$.  A primary cause of such detonations is that the shape of the
initially stationary burning front is distorted due to the effects of
the Rayleigh-Taylor (RT) instability (e.g., \citealt{bell2004}) and
convective turbulence in the surrounding fluid.
In Appendix \ref{s:zeldovich} we argue that the mixing of
fuel and ash due the convective motions present in RIAFs is indeed sufficient
for a detonation to occur via the Zel'dovich gradient mechanism \citep{Zeldovich+70} 
if $\Psi$ has the appropriate magnitude. 

If $\Psi \ll \Psi_{\rm crit}$, then large scale detonations do not develop and
accretion remains relatively steady across the burning front. 
During this `quiescent' mode of NuDAF evolution, some fraction of the
ash behind the burning front is accreted, while the rest is ejected into an
outflow along the polar axis.  Even when a detonation does not occur, NuDAFs
produce a more powerful outflow than in otherwise identical RIAFs without
nuclear burning. The fraction of the material which is gravitationally unbound
versus that being accreted is found to increase with $\Psi$ (\S\ref{s:quiescent},
Fig.~\ref{f:mdot_lowat}).  

\subsection{Detonations}
\label{s:detonations}

We begin by elaborating on the evolution of torii which undergo detonation,
basing our discussion on the high-resolution model COq050\_HR (Figure \ref{f:nudaf_truecolor}).  
The angle-averaged thermodynamic quantities in the disk as a function of radius
are shown in Figure~\ref{f:rhoT_timescales} at different times.
Also shown are various timescales in the disk as a function of radius, calculated at a
time $t = 1.673t_{\rm orb}$, corresponding to the onset of the final detonation.  

Because the sound crossing time is everywhere shorter than the viscous heating
time, pressure equilibrium rapidly establishes a temperature profile that
is nearly virial
($T \propto r^{-1}$) in regions of the disk where neutrino
cooling is negligible.  This profile becomes shallower than 
$r^{-1}$ interior to the burning radius, because neutrino cooling becomes
important and offsets viscous heating.  The density also obeys a power-law
radial profile (for $\rho > \rho_0$; eq.~[\ref{eq:density_cutoff}]), with a
slope that decreases with time as matter flows inwards.  Such a profile
steeping is expected since (at least in regular RIAFs) one expects $\rho
\propto r^{-1/2}$ once steady-state accretion is reached on a timescale $t \sim
t_{\rm visc}$.  

The inner edge of the torus ignites first, starting interior to the burning
radius $r_{\rm nuc}$ predicted by equation (\ref{eq:rnuc_definition}).  Burning
requires a significantly higher temperature (and hence smaller radii) than at
later times because initially the density is lower than 
$\rho_0$ (eq.~[\ref{eq:density_cutoff}]), below which nuclear reactions are
artificially suppressed.  If neutrino cooling is not included in our
calculations, then a global detonation is triggered at these very early times
if the value of $\Psi$ is sufficiently high
(model COq100\_nc).  Given that neutrino cooling is physically motivated and that
detonations via the Zel'dovich mechanism may be suppressed if the effects of
radiation pressure were properly included (Appendix \ref{s:zeldovich}), we
believe that such prompt detonations are 
unphysical, at least within
the parameter regime of this study.\footnote{During the tidal disruption of a
more massive O-Ne WD, nuclear burning may begin during the circularization
process itself.  A dynamical detonation appears more likely to occur in this
case.}  Verifying this will require simulations using a more detailed model of
the initial structure of the disk and a more physical EOS. 

When neutrino cooling is included or if $\Psi$ is lower, then a prompt
detonation is avoided and a steady-state burning front develops. The front settles at
the expected radius $r \approx r_{\rm nuc}$ within a few local dynamical times.
A detonation can still be triggered at later times as
the result of turbulence generated by fluid instabilities if $\Psi$ is
sufficiently high.  The RT instability is somewhat suppressed by neutrino
cooling when the value of $\Psi$
is sufficiently large. At low $\Psi$, or in the absence of cooling,
the burning front is noticeably distorted.
Turbulence should be a ubiquitous
feature of a more realistic MHD model since it can be generated directly by the
MRI, rather than originating second-hand from convective 
instabilities\footnote{Note however that in 3D, poloidal Rayleigh-Taylor mixing
can be affected by the Kelvin-Helmholtz instability in the orbital direction, 
potentially leading to suppressed growth (e.g., \citealt{chandrasekhar1961}).}. 

Figure~\ref{f:nudaf_ignition} shows a few snapshots of the region localized
around the burning front in model COq050\_HR, illustrating how turbulence
triggers a detonation.  
Between $1.65$ and $1.673$ orbits, a turbulent eddy near
the disk midplane mixes some of the hot downstream ash with the `cold' upstream
fuel, creating a hot spot just upstream of the burning front.  If mixing of ash
and fuel is such that that the inverse gradient in the induction timescale
$|\nabla t_{\rm ind}|^{-1}$ across the eddy exceeds the local sound speed,
then a localized burning front can transition into a detonation via the
so-called Zel'dovich grandient mechanism (\citealt{Zeldovich+70}; see Appendix
\ref{s:zeldovich}).  We believe this provides a reasonable explanation for the
delayed detonations observed in our simulations.  

Figure~\ref{f:spot_profiles} shows the mass fractions of fuel and ash across
the hot spot, confirming that mixing is indeed involved in raising the fuel
temperature.  Also shown are the sound speed and the inverse of the induction
time gradient across the eddy.  The latter peaks at the hot spot at a value
within a factor of $\sim 2$ of the sound speed, roughly consistent with the
Zel'dovich threshold for a spontaneous detonation. 
  
\begin{figure}
\includegraphics*[width=\columnwidth]{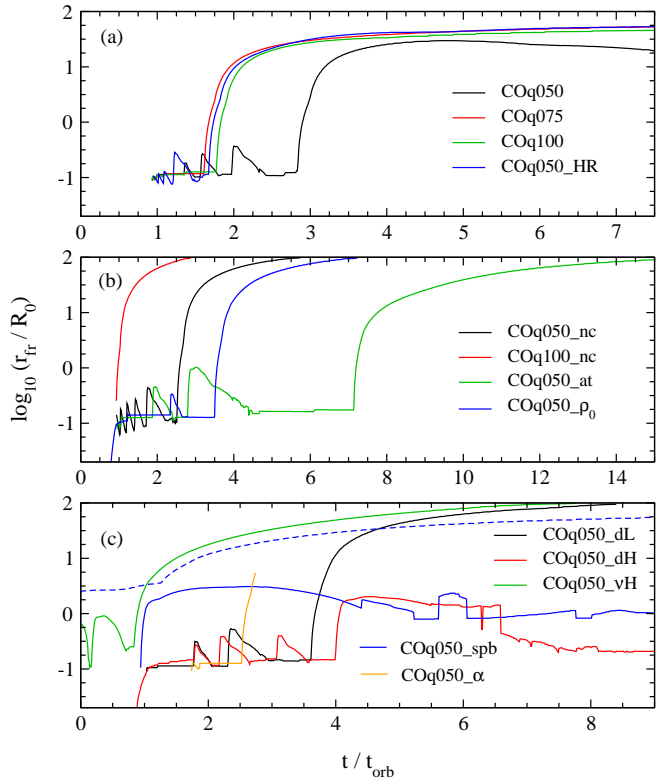}
\caption{Radius of the outer edge of the burning front at the disk equator
$r_{\rm fr}$ as a function of time, shown for several models that undergo
explosions.  The radius is calculated by 
searching for the outermost point where the ash mass fraction is equal to 0.25.  
Panel (a) shows models with high atmospheric density, while panels (b) and (c)
show models that vary different parameters around the marginal model
with low atmospheric density and neutrino cooling COq050\_at.
Note that the sawtooth shape of
some curves signals multiple failed detonations in cases where $\Psi$ is close
to the critical value $\Psi_{\rm crit} \simeq 1$ required for a successful
large-scale detonation. The burning radius is 
$r_{\rm nuc} \sim 0.1R_0$ for the models shown (Table~\ref{t:results}).
The dashed curve in model COq050\_spb in panel (c)
denotes the outer edge of the torus; in this model the leading shock
decouples from the burning front and stalls at $r\sim 50R_0$. 
}
\label{f:rfront_explode}
\end{figure}

\begin{figure}
\includegraphics*[width=\columnwidth]{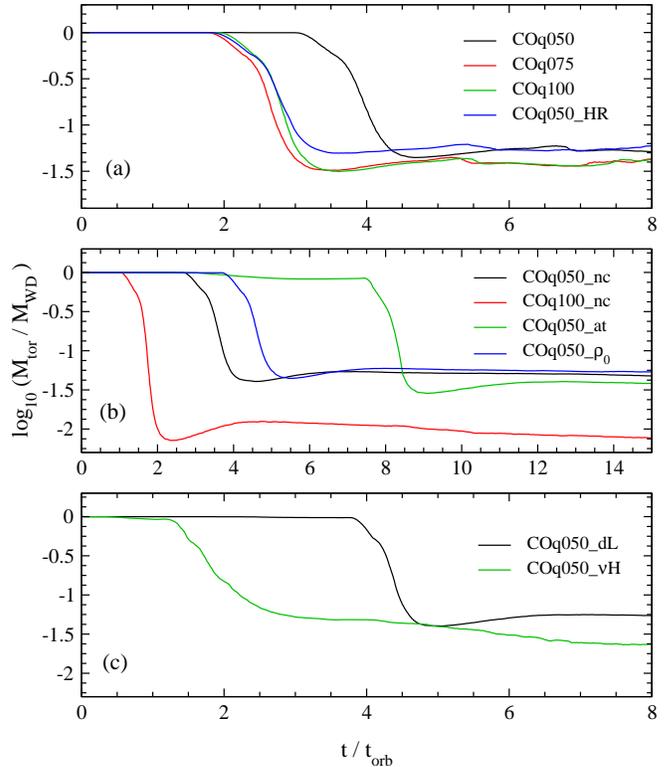}
\caption{Ratio of the mass in the torus $M_{\rm tor}$ to the
initial WD mass as a function of time, for models that 
explode (Figure~\ref{f:rfront_explode}).
The torus is defined as including all material with density $\rho \ge 10^{-3}\rho_{\rm max}$. 
Non-exploding torii are not shown, as the decrease in mass is negligible
on the same vertical scale.
After explosion, the ash torus
resumes accreting onto the central compact object, but at a reduced rate
compared to disks which do not detonate (see Figure~\ref{f:mdot_lowat}).
}
\label{f:disk_mass_time}
\end{figure}

We now discuss the fate of detonations when parameters of the system are varied.
The discussion is centered on variations in $\Psi$ (eq.~[\ref{eq:psi_def}]),
as it directly probes the dynamical effect of nuclear reactions on the disk.
For each model, the value of this parameter shown in Table~\ref{t:results} is measured by first
finding $r_{\rm nuc}$ just before explosion, and then computing the ratio
of $\varepsilon_{\rm nuc}$ to the angle-averaged and mass-weigthed enthalpy at $r=r_{\rm nuc}$.
For non-exploding models, a time average value is taken.
From equations~(\ref{eq:psi_pp}) and (\ref{eq:rnuc_definition}) one can see that the
(time-dependent) radial dependence of the temperature and density on radius
have the most influence on the time variation of $\Psi$ in a given model.
Nonetheless, the values obtained are generally close to what can be
inferred from the initial properties of the disk.

Depending on the magnitude of $\Psi$ relative to $\Psi_{\rm
crit}$, the hot spots generated by turbulence can produce three
different outcomes.  For $\Psi < \Psi_{\rm crit}$, detonations remain localized
and die out after propagating a distance comparable to the local radius.  
Multiple such \emph{microexplosions} occur as the density of the
inner accretion flow increases with time (Fig.~\ref{f:nudaf_ignition}).
Although some of these temporarily enhance the 
thickness of the disk,
accretion continues relatively steadily and
the burning front remains stationary at $r \approx r_{\rm nuc}$.

\begin{deluxetable*}{lccccccccccc}
\tablecaption{Summary of Results\label{t:results}}
\tablewidth{0pt}
\tablehead{
\colhead{Model}&
\colhead{$r_{\rm nuc}$\tablenotemark{a}} &
\colhead{$\Psi$\tablenotemark{b}} &
\colhead{$t_{\rm exp}$\tablenotemark{c}} &
\colhead{$v_{\rm exp}$\tablenotemark{d}} &
\colhead{$M_{\rm rem}$\tablenotemark{e}} &
\colhead{$t_{\rm ave}$\tablenotemark{f}} &
\colhead{$\langle \dot{M}_{\rm in}\rangle$\tablenotemark{g}} &
\colhead{$\langle \dot{M}_{\rm out}\rangle$\tablenotemark{h}} &
\colhead{$\langle X_{\rm a,out}\rangle$\tablenotemark{i}} &
\colhead{$\langle \dot{M}_{\rm unb}\rangle$\tablenotemark{j}} & 
\colhead{Notes\tablenotemark{k}}\\
 & $(R_0)$ &  & $(t_{\rm orb})$ & $(v_{\rm k0})$ & $(M_{\rm WD})$ & $(t_{\rm orb})$ & 
 \multicolumn{2}{c}{$(M_{\rm WD}/t_{\rm orb})$} & & $(M_{\rm WD}/t_{\rm orb})$ & 
}
\startdata
COq000         & \nodata & 0.00  & \nodata & \nodata & \nodata     & 15-20 & $10^{-2.6}$  & $10^{-2.2}$ & 0           & $10^{-4.3}$ & q\\
COq010         & 0.26    & 0.14  & \nodata & \nodata & \nodata     & 15-20 & $10^{-2.7}$  & $10^{-2.1}$ & $10^{-1.6}$ & $10^{-4.2}$ & q\\
COq025         & 0.21    & 0.31  & \nodata & \nodata & \nodata     & 15-20 & $10^{-3.1}$  & $10^{-1.9}$ & 0.2         & $10^{-3.8}$ & q\\
COq050         & 0.12    & 0.60  &  2.83   & \nodata & $10^{-1.2}$ & 8-10 & $10^{-3.7}$  & \nodata     & \nodata     & \nodata     & d,m,s\\
COq075         & 0.12    & 0.89  &  1.61   & \nodata & $10^{-1.4}$ & 8-10 & $10^{-3.8}$  & \nodata     & \nodata     & \nodata     & d,s \\
COq100         & 0.13    & 1.20  &  1.77   & \nodata & $10^{-1.4}$ & 8-10 & $10^{-3.8}$  & \nodata     & \nodata     & \nodata     & d,s  \\
\noalign{\smallskip}
COq000\_nc     & \nodata & 0.00    & \nodata & \nodata & \nodata     & 11-16 & $10^{-2.9}$ & $10^{-2.1}$ & 0       & $10^{-4.5}$ & q\\
COq010\_nc     & 0.23    & 0.13    & \nodata & \nodata & \nodata     & 11-16 & $10^{-3.1}$ & $10^{-2.1}$ & 0.1     & $10^{-4.5}$ & q\\
COq025\_nc     & 0.18    & 0.28    & \nodata & \nodata & \nodata     & 11-16 & $10^{-3.9}$ & $10^{-2.0}$ & 0.2     & $10^{-4.0}$ & q\\
COq050\_nc     & 0.10    & 0.48    &  2.51   & 2       & $10^{-1.3}$ & 11-16 & $10^{-4.3}$ & \nodata     & \nodata & \nodata     & d,m\\
COq100\_nc     & \nodata & \nodata &  0.93   & 5       & $10^{-2.1}$ & 11-16 & $10^{-5.1}$ & \nodata     & \nodata & \nodata     & p  \\
\noalign{\smallskip}
COq050\_HR      & 0.11 & 0.60  &  1.67   & \nodata & $10^{-1.2}$ & 8-10   & $10^{-3.7}$ & \nodata     & \nodata & \nodata     & d,m\\
COq050\_at      & 0.17 & 0.60  &  7.13   & 1       & $10^{-1.4}$ & 11-16  & $10^{-3.7}$ & \nodata     & \nodata & \nodata     & d,m\\
COq050\_$\rho_0$& 0.13 & 0.55  &  3.49   & 2       & $10^{-1.3}$ & 11-16  & $10^{-3.6}$ & \nodata     & \nodata & \nodata     & d,m\\
COq050\_dL      & 0.15 & 0.64  &  3.61   & 1       & $10^{-1.3}$ & 8-13   & $10^{-3.6}$ & \nodata     & \nodata & \nodata     & d,m\\
COq050\_dH      & 0.15 & 0.70  & \nodata & \nodata & 0.7 & 12-17 & $10^{-4.9}$ & $10^{-1.5}$ & $10^{-1.6}$    & $10^{-2.8}$  & q,m\\
COq050\_$\nu$H  & 0.25 & 0.63  &  0.83   & 1       & $10^{-1.6}$ & 6-8    & $10^{-2.9}$ & \nodata     & \nodata & \nodata     & d,m\\
COq050\_spb     & \nodata & \nodata  &  0.94   & \nodata & $0.19$      & 8-10   & $10^{-4.6}$ & \nodata     & \nodata & \nodata     & p,s\\
COq050\_ss      & 0.13 & 0.63  &  2.52   & \nodata & \nodata     & \nodata & \nodata     & \nodata     & \nodata & \nodata     & d,m\\
\noalign{\smallskip}
CO\_f1  & 0.21  & 0.91   & \nodata & \nodata & 0.87         & 11-16  & $10^{-4.2}$ & $10^{-1.8}$ & 0.3     & $10^{-2.0}$ & q\\
CO\_f2  & 0.11  & 1.00   &  0.56   & 1       & $10^{-1.9}$  & 9-11   & $10^{-3.4}$  & \nodata    & \nodata & \nodata     & d\\
HE\_00  & \nodata & \nodata & \nodata & \nodata & \nodata   & 1-3 & $10^{-2.2}$  & $10^{-0.9}$ & \nodata     & $10^{-1.7}$ & q\\
HE\_f1  & \nodata & \nodata & \nodata & \nodata & \nodata   & 1-3 & $10^{-2.2}$  & $10^{-0.9}$ & $10^{-5.5}$ & $10^{-1.7}$ & q\\
\enddata
\tablenotetext{a}{Radius of the steady-state burning front. For delayed explosions, measured just before the last detonation
		  sets in; time-averaged value for quiescent models. Not measured in prompt explosions. 
		  Ill-defined for He models}
\tablenotetext{b}{Ratio of $\varepsilon_{\rm nuc}$ to the enthalpy (angle-averaged and mass-weighted) 
		  at the burning radius (eq.~[\ref{eq:psi_def}]), right before explosion or time-averaged if non-exploding.}
\tablenotetext{c}{Time of the final detonation (exploding models only).}
\tablenotetext{d}{Expansion velocity at the time when shock reaches the domain boundary ($100R_0$), in units 
                  of $(GM_{\rm c}/R_0)^{1/2}$ (exploding models with low $\rho_\infty$ only).}
\tablenotetext{e}{Mass of the remnant torus (only applicable to exploding models).}
\tablenotetext{f}{Time range for average properties (quiescent models) or remnant mass (exploding models) calculation.}
\tablenotetext{g}{Time-averaged net accretion rate through the inner boundary.}
\tablenotetext{h}{Time-averaged net mass-loss rate at $r=4R_0$.}
\tablenotetext{i}{Ratio of net time-averaged mass-loss in nuclear ash to total ejecta, evaluated at $r=4R_0$.}
\tablenotetext{j}{Time-averaged mass-loss rate at $r=4R_0$, including unbound material only.}
\tablenotetext{k}{Outcomes: quiescent accretion (q), delayed explosion (d), 
                            multiple detonations before explosion (m), 
			    prompt explosion (p), ejecta stalls (s).}
\end{deluxetable*}

When $\Psi$ is in the vicinity of $\Psi_{\rm crit}$, hot spots can trigger large scale
detonations, but many of them die out as they propagate outwards through the
disk.  Because burning occurs at a radius $r_{\rm nuc}$ interior to where the
density of the torus peaks ($r \sim R_0$), such detonations are 
required to propagate against the density gradient. 
The weakening of the shock leads to a decrease in the resulting overpressure,
decreasing the burning time behind the shock and hence the energy released over
the region where the phase velocity of burning is supersonic. For slow enough burning,
the shock degenerates into a pressure wave and damps \citep{Niemeyer&Woosley97}.
Models with $\Psi \sim
\Psi_{\rm crit}$ are thus marginal in that they experience multiple failed
detonations before a final explosion.
Success is enabled in some cases by the changing radial
density gradient, which decreases as the inner torus viscously evolves
(Fig.~\ref{f:rhoT_timescales}).  Since the probability of a successful
detonation increases with time, one must follow the evolution 
for a sufficiently long time
before establishing
whether a marginal model will ultimately succeed or fail to explode.  

For large $\Psi \gg \Psi_{\rm crit}$, the energy released by even the first
detonation is sufficient to pass the point of maximum density.  
Our parametric sequence of simulations indicate that this critical value $\Psi_{\rm crit}
\approx 1$ appears to be relatively insensitive to 
whether neutrino cooling is
included or not.  The final column of Table \ref{t:results} summarizes the outcome of
each simulation. 

Figure \ref{f:rfront_explode} shows the time evolution of the burning front
radius $r_{\rm fr}$ at the disk midplane in several models which ultimately 
result in explosions. 
When $\Psi$ is large (such as in COq100), then the burning front remains near
$r_{\rm nuc}$ until the first 
detonation occurs.  The sawtooth shape of the burning front evolution
for `marginal' models (such as COq050) 
signals multiple failed detonations before the final successful explosion.  The top
and middle panels of Figure \ref{f:rfront_explode} compare $r_{\rm fr}(t)$ in
models with and without neutrino cooling. Changing the resolution does not
alter the qualitative outcome as inferred from comparing models COq050 and COq050\_HR.
The higher resolution model explodes earlier while still undergoing multiple detonations.

Variations around the marginal model with low atmospheric density (COq050\_at)
show that a lower thermal content and faster accretion are more conducive to
detonations. Models COq050\_dL and COq050\_$\nu$H both explode earlier than
the marginal model. A lower cutoff density $\rho_0$ also causes the disk
to explode earlier. Note however that the value of $\Psi$ is remarkably close
to $0.6$ for all the exploding models that use the power-law burning rate. 
In contrast, model COq050\_dH does not undergo a global detonation. Counterintuitively,
this model has a higher temperature normalization, but the detonation fails to sweep
through the whole disk. We surmise that the larger extent of the disk plays a role
in this failure. 

\begin{figure}
\includegraphics*[width=\columnwidth]{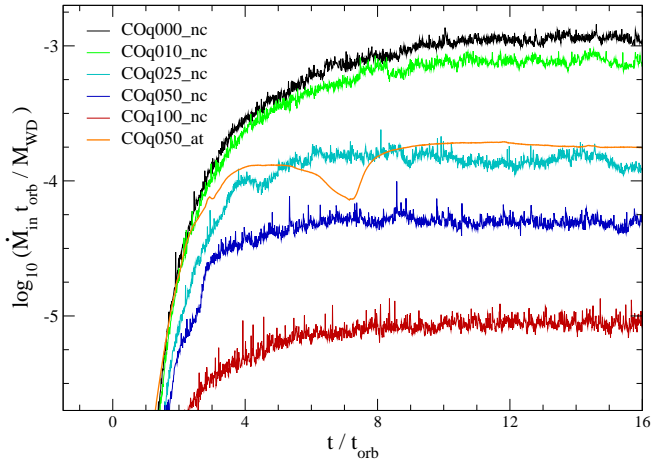}
\caption{Mass accretion rate $\dot{M}_{\rm in}$ at the inner boundary ($r_{\rm
in} = 0.01R_0$) as a function of time, shown for models with low atmospheric
density.  All models shown (except for COq050\_at) neglect neutrino cooling.
Note that as Q [$\Psi$] is increased from 0 to $Q_{0}$ [1.8], the peak
accretion rate decreases by over two orders of magnitude.  
The decrease in $\dot{M}_{\rm in}$ 
arises either as the result of a large-scale detonation
(when $\Psi > \Psi_{\rm crit}$) or due to enhanced convection and polar outflows (when
$\Psi < \Psi_{\rm crit}$; see Fig.~\ref{f:mass_flux_machrms}).}
\label{f:mdot_lowat}
\end{figure}

The functional form of the viscous stress does not appear to
greatly influence the outcome, so long as the strength is comparable.
Models COq050\_$\alpha$ and COq050\_spb probe the effects of using the
viscosity prescriptions in equations~(\ref{eq:viscosity_alpha}) and (\ref{eq:viscosity_spb}).
The amplitude is chosen so that the inner edge of the torus reaches the inner boundary
at around the same time as in model COq050\_at.
Whereas model COq050\_$\alpha$ undergoes a vigorous explosion (so much so that
the model crashes after 3 orbits), model COq050\_spb instead develops a prompt
explosion that fizzles because the leading shock and the burning front decouple
from one another. 

Models CO\_f1 and CO\_f2 employ the full functional form of the
$^{12}$C($^{12}$C,$\gamma$)$^{24}$Mg reaction, and assume the full energy
release ($Q=Q_0$).  They differ in their thermal content and strength of the
viscosity. The model with higher thermal content and lower viscosity does not
explode (CO\_f1), whereas the other (CO\_f2) does undergo a large scale
detonation.  
Because a reasonable variation in the model parameters results in
a qualitative change in the predicted outcome, this implies that the true
evolution is sensitive to the details of the accretion physics.  Determining
whether disk detonation is indeed a robust outcome of C-O WD-NS mergers in Nature
will thus require simulations which include both a more physical EOS and
account for other details, such as the true MHD nature of the disk turbulence.

Finally, we find that the He disk model with the triple-alpha reaction (HE\_f1) fails to
detonate, in part because the burning begins at a radius ($\sim 0.02R_0$) where $\Psi$ is small.
This is due to the weaker temperature dependence of the reaction rate, which leads to a burning
front that is more spread out in radius than in the C-O WD case.
In fact, the mass fraction of carbon achieves a maximum of $3\times 10^{-4}$
during the simulated period (4 orbits at $R_0$), making the concept of a burning radius ill-defined. 
%

\begin{figure}
\includegraphics*[width=\columnwidth]{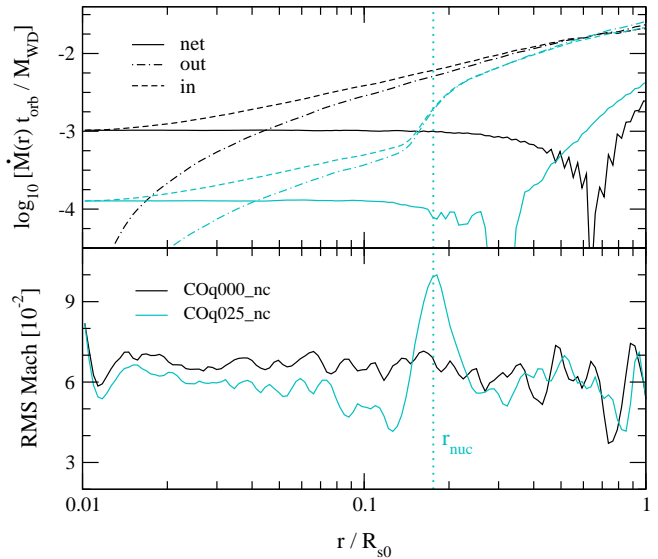}
\caption{\emph{Top:} Comparison of time-averaged mass fluxes as a function of 
radius in non-exploding models COq000\_nc (no burning, black curves) and 
COq025\_nc ($Q/Q_0 = 0.25$, cyan curves). 
Outflow, inflow, and net fluxes are shown as dot-dashed, dashed, and solid curves,
respectively. The vertical dotted line corresponds to the time-averaged burning
radius $r_{\rm nuc}$. \emph{Bottom:} Root-mean-square Mach number as a function of
radius in the disk midplane for the same set of models. The energy deposition from
nuclear burning causes a localized increase in the turbulent kinetic energy, decreasing
the net mass flux through the disk inside $r_{\rm nuc}$ (compare with
Figure~\ref{f:mdot_lowat}).}
\label{f:mass_flux_machrms}
\end{figure}

Table~\ref{t:results} gives the time of the final detonation for each exploding model and
the shock velocity once it reaches the outer simulation boundary at $r=100R_0$.
In models with low atmospheric density ($\rho_{\infty}= 10^{-7.7}\rho_{\rm
max}$) the velocity asymptotes to a nearly constant value of order of the
Keplerian speed near the radius of the original torus $v_{\rm k0} = (GM_{\rm
c}/R_0)^{1/2}$ ($\sim 3000$~km~s$^{-1}$ in our fiducial C-O WD models).  By
contrast, in models with high background density ($\rho_{\infty} =
10^{-5.7}\rho_{\rm max}$) the velocity decreases and the radius of the burning
front comes to an effective halt.  This outcome is an artifact of the high
inertia of the atmosphere, as is clear by comparing the evolution of models
COq050 and COq050\_at, which differ only in the value of $\rho_{\infty}$.  In
order to accurately predict the asymptotic properties of outflows from the
disk, and hence their resulting observational signatures, one must use a
sufficiently low atmospheric density for which the shock motion is converged.
We have not undertaken such a convergence study here, since our primary goal is
understanding the basic dynamics of the disk on smaller scales.  Also note that
the density suppression of the burning rate (eq.~[\ref{eq:density_cutoff}])
ensures that no burning takes place after the detonation has swept the initial
torus material. 

Disks that explode leave behind a remnant torus composed primarily of ash.
Figure~\ref{f:disk_mass_time} shows the time evolution of the torus mass
$M_{\rm tor}$, defined as that enclosed within the density contour $\rho
\ge 10^{-3}\rho_{\rm max}$, in several disk models.  
The remnant disk after explosion contains between 1\% and 10\% of
the initial WD mass, depending on the value of $\Psi$ and on whether neutrino
cooling is included.  The existence of such a disk implies that the accretion
rate onto the central compact object does not immediately fall to zero in disks
that explode (see also Figure \ref{f:mdot_lowat}).      

\begin{figure*}
\includegraphics*[width=\textwidth]{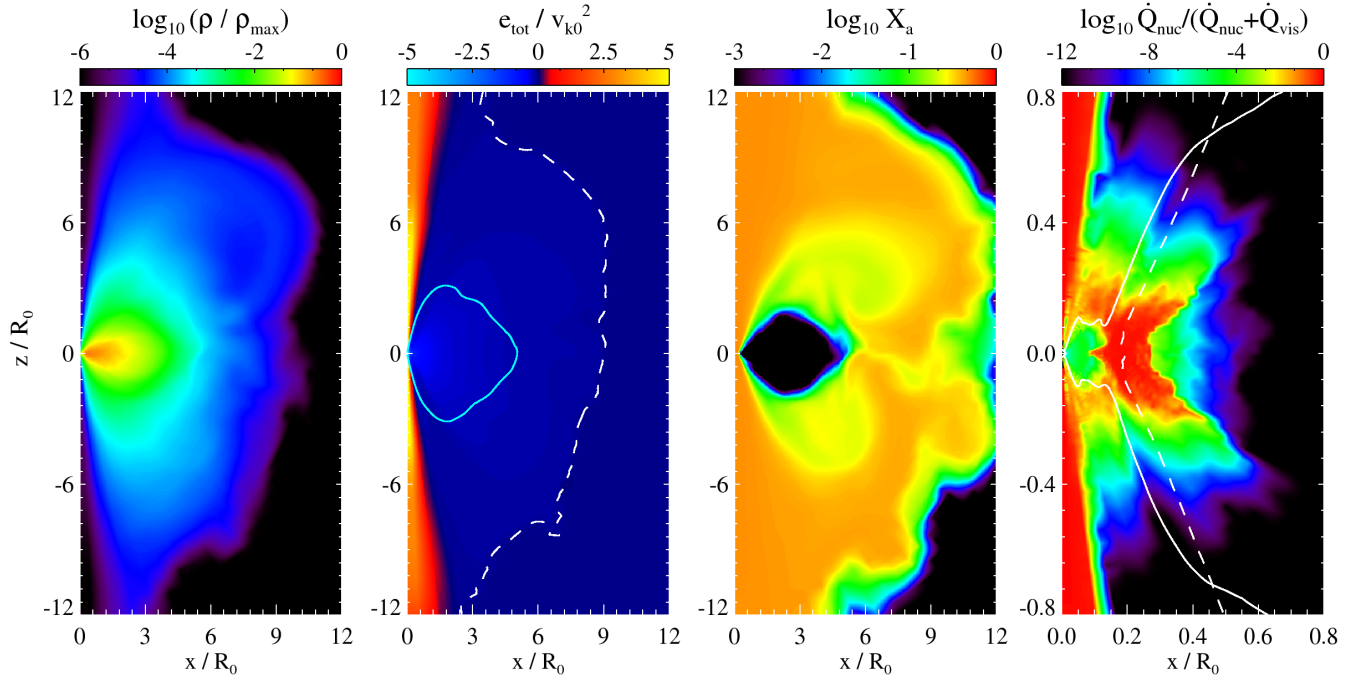}
\caption{Time-averaged properties of the quiescently accreting NuDAF model
COq025\_nc, averaged between orbits 11 and 16.  Quantities displayed (from left
to right) include the density $\rho$; total energy $e_{\rm tot}$ (thermal +
kinetic - gravitational); mass fraction of ash $X_{\rm a}$; and fraction of
the total heating $\dot{Q}_{\rm tot} = \dot{Q}_{\rm nuc} + \dot{Q}_{\rm vis}$
contributed by nuclear heating.  The light blue contour in the energy panel
corresponds to $\rho/\rho_{\rm max}=10^{-3}$, marking the approximate
surface of the torus. The dashed white contour shows the region where the
mass fraction of atmospheric material is more than $0.1\%$ by mass (the atmosphere
is very unbound by construction; we set its energy to zero in this plot to
maximize contrast between regions dominated by torus material).
The solid white contour on the heating ratio
panel corresponds to $\dot{Q}_{\rm visc} = 10^{-4}v_{k0}^3/R_0$, while the
dashed contour denotes the burning front ($X_{\rm a} = 0.25$). Note that both heating
terms are suppressed for $\rho < 10^{-3}\rho_{\rm max}$
(eq.~[\ref{eq:density_cutoff}]), hence there is negligible heating inside the
polar funnel.  Also note the different spatial scale of the rightmost panel.}
\label{f:nudaf_individual}
\end{figure*}

\begin{figure}
\includegraphics*[width=\columnwidth]{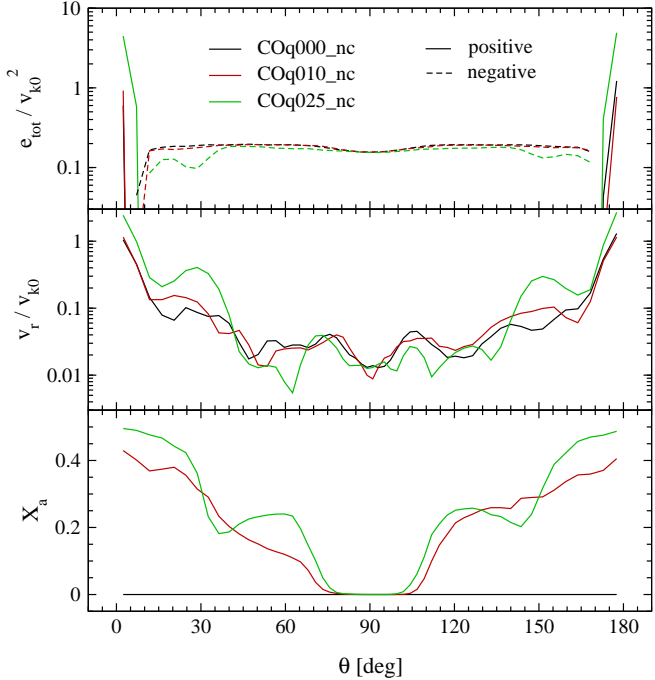}
\caption{Time-averaged quantities at $r=4R_0$ as a function of polar angle, for three
quiescent models that differ in the strength of nuclear burning. Shown are (a) the total energy,
(b) radial velocity, and (c) mass fraction of nuclear ash.}
\label{f:angular_quantities}
\end{figure}

\begin{figure*}
\includegraphics*[width=\textwidth]{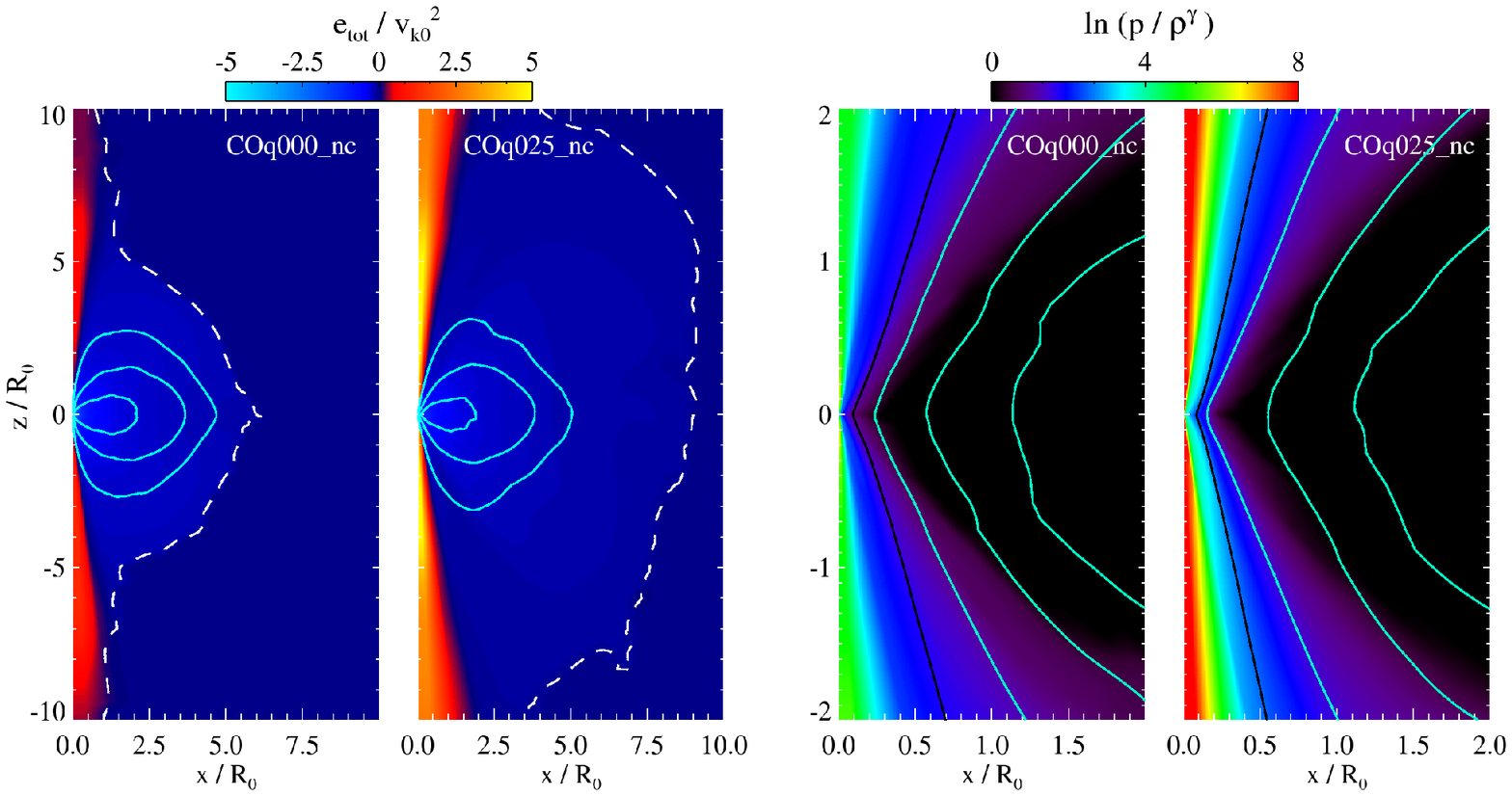}
\caption{Comparison of time-averaged properties between a model with nuclear
burning (COq025\_nc) and one without (COq000\_nc).  The average is taken
between orbits 11 and 16.  The left two panels show the total energy, with
solid contours corresponding to densities $\rho/\rho_{\rm
max}=\{10^{-3},10^{-2},10^{-1}\}$. White dashed contours mark the surface
where the atmosphere is $0.1\%$ of the fluid by mass (outside of which
we plot the total energy as zero, to maximize contrast).
The right two panels show a function
proportional to the entropy of the gas, zoomed in on the region surrounding the
burning front.  Contours correspond to angular momenta 
$\ell_z\, =\{0.25,0.4,0.65,0.9\}  (GM_{\rm c}R_0)^{1/2}$.  
The higher entropy of the polar funnel in
the case of nuclear burning indicates excess heating.  This is seen
in the difference between the time-average mass loss rate in unbound material,
which is $\langle \dot{M}_{\rm unb}\rangle = 10^{-6}M_{\odot}$~s$^{-1}$ and 
$3\times 10^{-7}M_{\odot}$~s$^{-1}$ in the cases with and
without nuclear heating, respectively.}
\label{f:nudaf_comparison}
\end{figure*}

\subsection{Quiescent Disks and Explosion Remnants}
\label{s:quiescent}

Although small-scale runaway burning and localized microexplosions may occur
for $\Psi < \Psi_{\rm crit}$, the disk as a whole never detonates.  The
properties of disks with nuclear heating are nevertheless qualitatively
different from those without burning.
In this section we discuss these NuDAFs which undergo quiescent accretion.

Figure~\ref{f:mdot_lowat} shows the time evolution of the net mass accretion
rate at the inner boundary $\dot{M}_{\rm in}$ for the sequence of C-O torii
without neutrino cooling, for various values $\Psi$.  This sequence was
chosen since its low atmospheric density $\rho_\infty$ allows us to better
study the properties of disk outflows. 
The asymptotic accretion rate monotonically decreases with increasing $\Psi$,
with $\dot{M}_{\rm in}$ dropping
by more than two orders of magnitude as $\Psi$ increases from 0 to $\sim 1$.  For
large values of $\Psi > \Psi_{\rm crit}$, this decrease in the accretion rate
is the result of the mass loss caused by the disk detonation, which leaves only
a small remnant torus (Fig.~\ref{f:disk_mass_time}).  However, even for disks
that do not explode, the mass accretion rate is
substantially reduced by the effects of nuclear heating (e.g., model
COq025\_nc).  

The decrease in the net accretion rate in quiescent NuDAFs results from
more vigorous convection, driven by the enhanced heating from nuclear reactions.
This is illustrated in Figure~\ref{f:mass_flux_machrms}, which shows the
time- and angle-averaged mass fluxes as a function of radius for models with and without
nuclear burning. A localized enhancement in the turbulent kinetic energy (as measured
by the root-mean-square Mach number) around the time-averaged burning radius coincides
with a sharp drop in the mass fluxes in model COq25\_nc.

For comparison, Figure \ref{f:mdot_lowat} also includes a low-$\rho_{\infty}$
model {\it with} neutrino cooling included (COq050\_at). The net accretion
rate for this model is a smoother function of time than the otherwise
identical model without cooling (COq050\_nc), due to the near cancellation
of viscous heating by neutrino cooling (c.f.~Figure~\ref{f:rhoT_timescales}, bottom panel), 
which suppresses convection. This
is also responsible for the larger magnitude of the net mass flux.

In order to better understand the properties of disk outflows from quiescent
NuDAFs, we perform time-averages of the flow over several orbits once they have reached
a quasi-steady state (c.f. \citealt{stone1999}).
Figure~\ref{f:nudaf_individual} shows two-dimensional maps of density, total energy,
ash mass fraction, and heating strength for model COq025\_nc, 
averaged between orbits 11 and 16.
Material is unbound when its total energy
\begin{equation}
e_{\rm tot} = \frac{1}{2}\left[v_r^2+v_\theta^2 + \frac{\ell_z^2}{(r\sin\theta)^2}\right]
              + e_{\rm int} - \frac{GM_{\rm c}}{r},
\end{equation}
is positive. Figure~\ref{f:nudaf_individual} shows that disk material satisfying
this condition is confined to within a narrow funnel $\sim 10^\circ$ from the polar axis.
This unbound flow is composed primarily of nuclear ash and inert species. The distribution
of nuclear ash is wider than this funnel, however. Figure~\ref{f:nudaf_individual}
also shows that a significant fraction of the outflowing ash is bound, circulating
above the disk and returning to the midplane at distances larger than the outer
radius of the disk. The angular distribution of the total energy, radial velocity,
and ash for the quiescent models with no neutrino cooling are shown in Figure~\ref{f:angular_quantities}.
Note that the expansion velocity along the funnel can reach the Keplerian velocity  $v_{\rm k0}$
at the circularization radius, which for the fiducial C-O WD-NS merger is $\sim 3000$~km~s$^{-1}$.

The rightmost panel in Figure~\ref{f:nudaf_individual} shows the fraction of the total
heating in the disk contributed by nuclear burning.
Nuclear heating is important relative to
viscous heating in a region centered around $r_{\rm nuc}$ on the disk midplane.
Note that the time-averaged thickness of this region ($\sim r_{\rm nuc}$) is
larger than the instantaneous thickness of the burning region in the case of a
detonation (Fig.~\ref{f:spot_profiles}).  

Figure~\ref{f:nudaf_comparison} compares the time-averaged disk properties of
otherwise identical models with and without nuclear burning (COq025\_nc and
COq000\_nc, respectively).  The most significant effect of nuclear burning lies
in the properties of the polar funnel, which is 
more unbound than in the model without nuclear burning.
This difference also manifests in the entropy, which is higher inside the
funnel for disks with burning.  
A higher entropy indicates excess heating in
low temperature regions, as could result from additional heating above the
midplane due to the enhanced convection from nuclear burning
(Fig.~\ref{f:mass_flux_machrms}).
An entropy-increasing heating profile is
precisely the kind necessary to power an energetic outflow, as is well known in
other astrophysical contexts such as the solar wind.  

Nuclear burning does not greatly affect the structure of the disk
($\rho > \rho_{0}$; eq.~[\ref{eq:density_cutoff}]), although regions with lower
density are $\sim$50\% more extended in radius.  The presence of nuclear
burning also does not alter the fact that surfaces of constant entropy and
angular momentum are parallel, as is found in hydrodynamic simulations of RIAFs
(e.g., \citealt{stone1999}).  This indicates
that the disk is still marginally stable to convection, even when the energy
input by nuclear heating is included.

Table \ref{t:results} gives the time-averaged net
accretion rate $\langle \dot{M}_{\rm in}\rangle$ at the inner boundary, 
total outflow rate, as well as unbound outflow
at $r=4R_0$
in our quiescently-accreting models.  
Note that $\langle \dot{M}_{\rm unb}\rangle$ increases with $\Psi$.
In non-exploding models with parametric burning rate, the mass loss in unbound
material approaches a substantial fraction
of $\langle \dot{M}_{\rm in}\rangle$ as $\Psi \rightarrow \Psi_{\rm crit}$.
The fraction of ash in the unbound outflow also correlates with $\Psi$.

The decrease of the mass accretion rate with $\Psi$ allows the disk to survive for a time longer
than the characteristic viscous time. Model CO\_f1 is an extreme in this sense.
The mass-loss in unbound flows exceeds the accretion rate by two orders of magnitude.
Most of the WD material will be ejected over $\sim 100$ orbits, or
about $1.5$ hours as an unbound flow with a velocity $\sim 1000$~km~s$^{-1}$.
Converting to physical units, the mass loss rate in
unbound material can lie in the range $10^{-5}$ -- $10^{-3}$ $M_\sun$~s$^{-1}$
for the full range of models explored.

\section{Discussion}
\label{s:discussion}

\subsection{Implications for WD-NS/BH mergers}
\label{s:merger_discussion}

An obvious implication of our results is that the C-O torii created by WD-NS
can undergo a large-scale detonation and explosion.  
This explosive evolution
differs drastically from the steady-state model for
NuDAFs envisioned by \citetalias{M12},\footnote{\citetalias{M12} did show that his solutions were prone
to thermal instability due to the sensitive temperature dependence of nuclear
reactions.  Although one interpretation of our results is that such
instabilities manifest as a detonation, the \citetalias{M12} model only accounts for the
properties of the mean flow and not for the role of turbulence, which plays
such an essential role in the detonation process in our simulations.} although
the \citetalias{M12} model does provides a qualitative description of the quiescent mode of
accretion found to occur when $\Psi < \Psi_{\rm crit}$.  A similar conclusion
would likely apply to the case of a hybrid He-C-O WD because the reaction
$^{4}$He($^{16}$O,$\gamma$)$^{20}$Ne has a similar threshold $\Psi_{\rm crit}$
for detonation (Table \ref{table:psivalues}).  Given the ubiquity of turbulence
generated by the MRI in physical accretion disks and relative ease with which
the requisite conditions for a detonation are met (Appendix \ref{s:zeldovich}),
it is possible that detonation is a robust outcome of any disk with $\Psi
\gtrsim 1$ under the influence of a sufficiently temperature-sensitive nuclear
reaction.  Our picture of turbulence-generated
detonations in many ways resembles that invoked in `pulsation delayed
detonation' models for Type Ia SNe (e.g.,~\citealt{Khokhlov+97};
\citealt{Niemeyer&Woosley97}).\footnote{How the deflagration ignited near the
core of a Chandrasekhar-mass C-O WD transitions into a detonation remains a
major theoretical uncertainty in standard Type Ia supernova models.  Assuming
that the initial deflagration fails to unbind the WD, delayed detonation models
invoke a secondary detonation which results from the mixing of partially-burned
fuel and ash following the subsequent contraction of the WD.}

Although detonations appear to be a common feature of disk evolution when $\Psi
\gtrsim 1$, it is challenging to make a definitive statement about the ubiquity
of a detonation because there are hints that the outcome could depend on more
parameters than $\Psi$ alone.  For instance, our models CO\_f1 and CO\_f2 with full reaction
rates included have relatively minor differences in their parameters, yet one
explodes while the other does not.  The precise composition of the WD may also
matter since $\Psi$ is proportional to the fuel mass fraction $X_{\rm f}$.

Another important caveat is that our current EOS includes only ideal gas pressure,
despite the fact that radiation pressure is important at small radii and early
times in the disk evolution.  Since radiation pressure acts to reduce the
temperature at a given pressure, this implies that (1) a deeper potential well
is required for burning, thus reducing the value of $\Psi$; and (2) the likelihood
that runaway burning will occur is also reduced since the magnitude of
temperature fluctuations across the burning front are suppressed (Appendix
\ref{s:zeldovich}).  It is thus possible that radiation pressure could suppress
a detonation at early times in the torus evolution.  However, this should not
be the case once steady-state accretion is achieved on a timescale $\sim t_{\rm
visc}$ since then the density of the inner accretion flow is sufficient for gas
pressure to dominate.  The radially-decreasing density profile achieved on a
timescale $t \gtrsim t_{\rm visc}$ is also the most conducive to a sustaining
an outward-propagating detonation ($\S\ref{s:detonations}$).  This gives us
confidence that the threshold for detonation found in our simulations is not an
artifact of our EOS.

One consequence of a relatively prompt detonation is that a large fraction of
the WD is unbound, which substantially limits the mass ultimately accreted by
the central NS.  This makes it less likely that a black hole will be created
following a WD-NS merger (\citealt{Paschalidis+11}). A prompt explosion also 
limits the energy of a relativistic outflow (i.e.~ a `jet') originating from the vicinity of the
compact object (\citealt{King+07}) and its resulting high energy or radio
emission.  Even when a strong detonation occurs, however, the existence of a
remnant disk of bound material (Fig.~\ref{f:disk_mass_time}) indicates that a
small mass $\sim 10^{-2}M_{\odot}$
is still available to accrete.

Our highly idealized initial conditions are necessarily different from torii
that arise in self-consistent mergers. On the one hand, the disruption process is not instantaneous 
(e.g., \citealt{fryer1999}), thus the angular momentum distribution will not be constant.
Also, the thermalization of the rotational kinetic energy by shocks likely
leads to a non-uniform entropy distribution in the disk. If detonations 
early in the accretion phase turn out to be a prevalent outcome, then 
the initial form of the torus will be an important factor to consider when making
observational predictions. This would also imply that the angular momentum
transport process must be treated correctly. However, we do not expect
some of the more general results of this paper, such as the existence of a critical value
of $\Psi\sim 1$ for detonation or the dependence of quiescent outflows on $\Psi$, 
to depend fundamentally on the details of the initial condition.

In the case of a WD-BH merger, the larger mass of the central compact object
results in a lower value of $\Psi \propto M_{\rm c}^{-1}$, making it more
likely that accretion will instead occur in the quiescent regime ($\Psi <
\Psi_{\rm crit}$).  If non-explosive burning allows a larger fraction of the WD to
accrete, then the power of an associated high energy counterpart
(\citealt{fryer1999}) may be larger than in the WD-NS case.  One limitation on
the rate of such events is that a binary with a high mass ratio is required for
unstable
mass transfer to occur in the first place ($q \gtrsim 0.2-0.5$;
\citealt{Paschalidis+09}). The low mass black holes ($M_{\rm c} \lesssim 5
M_{\odot}$) thus required appear to be relatively rare among the population of
Galactic binaries (e.g.,~\citealt{Bailyn+98}; \citealt{Fryer+12}).

In contrast to C-O torii, He torii appear unlikely to explode, even
under conditions we find conducive to a detonation (Tables~\ref{t:models} and 
\ref{t:results}). Stability results
mostly from the weaker temperature dependence of the triple$-\alpha$ nuclear
reaction rate at high temperature, which makes the conditions for detonation
via turbulent mixing less likely to be satisfied (Appendix \ref{s:zeldovich}).
The torus density in our simulation is sufficiently low that a large fraction
of $^{4}$He may reach sufficiently high temperature to photodissociate before
burning into heavier elements.  Our conclusion that He torii are unlikely to
detonate is at odd with the suggestion of \citet{Schwab+12} that such
detonations might occur in disks created by WD-WD mergers.  One possible
difference in their case could be the importance of degeneracy pressure or the
higher densities achieved by compression due to the presence of a WD surface.

Another limitation of applying our current simulations directly to physical
mergers is that we include only a single reaction.  In some cases this could
affect our conclusions about the likelihood of a detonation.  Although 
pure helium torii appear to be stable, the products of He burning ($^{12}$C or
$^{16}$O) are themselves prone to explosive burning.  In C-O torii, on the
other hand, subsequent reactions (e.g.,~oxygen burning) will generally occur
much deeper in the potential well (lower $r_{\rm nuc}$ and hence lower $\Psi$;
Table \ref{table:psivalues}), such that the first reaction has the largest
dynamical impact.  Nevertheless, including a full reaction network could result
in a much richer evolutionary history, with, for instance, several potential
detonations occuring in succession as matter slowly sinks deeper into the
potential well and burns to increasing heavier elements.  This behavior may
qualitiatively resemble the late stages of unstable shell burning in massive
stars (e.g.,~\citealt{Arnett&Meakin11}; \citealt{Quataert&Shiode12}) or
pulsational pair instabilitity supernovae (e.g.,~\citealt{Woosley+07}).

\subsection{Supernova-Like Optical Transients}

\citetalias{M12} proposed that the ejecta from a WD-NS/BH merger could produce an optical
transient, powered by the radioactive decay of $^{56}$Ni synthesized in the
accretion disk.  He was motivated by the recent discovery of several Type I
supernovae which are dimmer and/or more rapidly evolving than normal SNe Ia or
Ib/c (e.g.,~\citealt{Li+03}; \citealt{Jha+06}; \citealt{Foley+09};
\citealt{Perets+10}; \citealt{Waldman+11}; \citealt{Kasliwal+11}).  The rapid
evolution and low luminosities of these events require both lower $^{56}$Ni
yield and lower total ejecta mass than values characteristic of normal SNe.  Some of these
events occur in locations far outside of their host galaxies
(\citealt{Perets+10}; \citealt{Kasliwal+11}), possibly consistent with the
location of a WD-NS merger if the NS was given a natal `kick'.

Based on a steady-state model of accretion following a WD/NS-BH merger, \citetalias{M12}
predicted an ejecta composed primarily of unburned C-O or He (depending on the
initial WD) along with a range of intermediate mass elements and a small
quantity of $^{56}$Ni.  Since our calculations in this paper include only a
single reaction, we cannot directly predict the Ni yield or precisely determine
the velocity structure of the ejecta.  Nevertheless, we can address whether
qualitative features of the outflows are consistent with those required to
produce a supernova-like transient.

In the case of quiescent disk evolution ($\S\ref{s:quiescent}$), the picture
that we find is qualitatively similar to that of \citetalias{M12}.  Nuclear reactions power
a quasi-steady outflow with an enhanced mass-loss rate over the case without
nuclear burning (Fig.~\ref{f:nudaf_comparison}).  If $\Psi < \Psi_{\rm crit}$
is satisfied for the first reaction activated at large radii in the disk, then
this condition will also be satisfied for subsequent reactions which release at
most a comparable amount of energy but occur deeper within the potential well.
The total mass of $^{56}$Ni ejected is indeed likely to be small $M_{\rm Ni}
\sim 10^{-3}-10^{-2}M_{\sun}$, because mass loss from the outer disk cuts off
the supply reaching smaller radii where the $^{56}$Ni produced.  Note, however,
that since our simulations show that unbound outflows from the disk primarily
originate from regions interior to the burning radius
(Figs.~\ref{f:nudaf_individual}, \ref{f:angular_quantities}), this suggests
that the \citetalias{M12} model may overestimate the fraction of unburned fuel in the
ejecta.

On the other hand, if the disk undergoes a global detonation
($\S\ref{s:detonations}$), then this picture is drastically altered.  In this
case the yield of intermediate mass elements and $^{56}$Ni will instead depend
on how far nuclear reactions proceed behind the detonation front.  Since the
average density of the outer disk is relatively low compared to that of the
original WD, the mass of $^{56}$Ni synthesized will almost certainly be much
less than in a normal Type Ia SNe (cf.~\citealt{Sim+10}).  How much Ni is
produced will depend on the time of detonation $t_{\rm det}$ relative to the
viscous time of the torus $t_{\rm visc}$ (eq.~[\ref{eq:tacc}]).  If $t_{\rm
det} \ll t_{\rm visc}$, then only a small fraction of the torus mass has spread
to small radii where the density is sufficiently high for $^{56}$Ni to be
produced, while if $t_{\rm det} \sim t_{\rm visc}$ then an larger fraction of
the shocked WD material will be processed to $^{56}$Ni.  Our current
simulations show that the former case ($t_{\rm det} \ll t_{\rm visc}$) is more
likely, but it is still possible that the latter case is more physical if
radiation pressure indeed suppresses early detonations.

\subsubsection{Connection to SN 2002cx-like Events?}

One type of SNe with characteristics seemingly compatible with those resulting from a
WD-NS/BH merger are the events prototyped by SN 2002cx and SN 2005bj 
(``SN 2002cx-like'' events; \citealt{Li+03};
\citealt{Jha+06}; \citealt{Phillips+07}; \citealt{Foley+09}).  This rare class
of Ia SNe are in part distinguished by the following features: (1) high
ionization spectrum at maximum light, and the presence of elements lighter than
the Fe group at all epochs, both indicating that the ejecta is extensively
mixed; (2) a low peak luminosity compared to that expected given its rate of
decline (i.e. not following the standard `Phillips relation'), indicating a
wide range of synthesized $^{56}$Ni mass, ranging from $M_{\rm Ni} \sim 3\times
10^{-3}-0.2M_{\odot}$; (3) low expansion velocities $\sim 3000-5000$ km
s$^{-1}$ (hence low kinetic energy) compared to a normal SN Ia ($\sim 10,000$
km s$^{-1}$); (4) permitted Fe II lines and continuum photospheric emission at
late times, indicating very low velocities $\lesssim 1,000$ km s$^{-1}$ for the
innermost ejecta; and (5) host galaxies with both active star formation
(\citealt{Foley+09}) as well as older stellar populations (\citealt{Foley+10b}). (6) An estimated occurence rate $\lesssim 10\%$ of that of normal SNe Ia (e.g.~\citealt{Phillips+07}).

Many of the above characteristics are qualitatively consistent with the
expected SN produced by a disk detonation following a WD-NS or WD-BH merger.
Given that a wide range in disk densities is expected depending on the mass
ratio of the binary, one likewise expects a wide range in the mass of $^{56}$Ni
and intermediate mass elements.  Extensive mixing also seems plausible given
the asymmetric nature of the explosion.  Low velocity of the inner ejecta is
also predicted, since the inner ejecta is substantially slowed by the
gravitational field of the central compact object (not present in a usual Type
Ia event; although see \citealt{jordan2012,kromer2012}).  
Even though 
the expected population of host galaxies is difficult to
predict with confidence, a mixture of both early and late-type galaxies is
likely (e.g.~\citealt{Belczynski+02}), given the distribution of gravitational wave inspiral times.  The lack
of spectroscopic evidence for unburned $^{12}$C in SN 2002cx-like events is
also compatible with a disk detonation model, though this observation is in
tension with alternative models invoking the pure deflagration of a
Chandrasekhar-mass WD (e.g., \citealt{Blinnikov+06}, \citealt{jordan2012}, \citealt{kromer2012}).  
The rate of WD-NS
mergers in the Milky Way is also estimated to be $\sim 10^{-4}$ yr$^{-1}$
(based on known population of tight WD-NS binaries;
\citealt{OShaughnessy&Kim10}), or approximately $\sim 3 \%$ of the SN Ia rate.
One potential problem with the disk detonation model is the continuum
polarization, which one would naively expect to be high given the large
asymmetry, yet at least in one case (2005hk) was observed to be small ($\sim
0.4\%$; \citealt{Chornock+06}).  

We will explore the properties of supernovae from WD-NS/BH mergers in greater
detail in future work using the properties of outflows from the torus including
a full $\alpha-$reaction network.  These will allow us to better test whether
WD-NS/BH mergers can be associated with 2002cx-like events or other classes of
subluminous Type I SNe.

\subsection{Implications for Collapsar Accretion Disks}

Another application of our results is to accretion disks formed by the collapse
of rotating stars, as in collapsar models for gamma-ray bursts (GRBs)
\citep{MacFadyen&Woosley99,lopez2009,Lindner+12}.  Depending on the
angular momentum profile of the star (\citealt{Woosley&Heger06}), a significant
fraction of the collapsing envelope may circularize at sufficiently large radii
$\gtrsim $ few $\times 10^{8}$ cm, where nuclear reactions such as
$^{4}$He($^{16}$O,$\gamma$)$^{20}$Ne are important.  Given the relatively low
mass of the black hole formed by the collapse of a Wolf-Rayet star, one expects
$\Psi \sim \Psi_{\rm crit}$ in collapsar disks, such that nuclear burning could
indeed impact the dynamics of the accretion flow.  If the effects of nuclear
burning produce large amplitude fluctuations in the central accretion rate, then
the viscous or dynamical timescale near the burning radius ($\sim$ seconds for
$r_{\rm nuc} \sim$ few $\times 10^{8}$ cm) could be imprinted on the
variability of GRB emission.  This could explain, for instance, why the power
density spectra of GRB light curves peaks at a frequency $\sim$ Hz
(\citealt{Beloborodov+00}) which is much lower than the $\sim$ kHz variability
characteristic of the innermost stable orbit.

Quiescent outflows, or periodic episodes of runaway burning, could also provide
a source of $^{56}$Ni, as is required to power the light curves of GRB
supernovae.  \citet{Bodenheimer&Woosley83} first showed that a sufficiently
rapidly rotating star could undergo explosive oxygen burning upon collapse, due
to heating caused by the centrifugal `hang up' of infalling material
(cf.~\citealt{MacFadyen&Woosley99}).  How much $^{56}$Ni is synthesized from
such explosive burning of stellar fuel, versus that produced within much hotter
outflows from the inner torus (\citealt{MacFadyen&Woosley99};
\citealt{Metzger+08}; \citealt{Milosavljevic+12}; \citealt{Surman+11}), has yet to be quantified.
Determining whether nuclear burning is in fact relevant to collapsar disks will
require global simulations of the collapsing star, including both nuclear
burning (as included here) and the possible influence of feedback from the
inner accretion flow (e.g., convection or a jet).  It will also require an
assessment of whether a stellar progenitor with a sufficiently large angular
momentum is physically realistic.

Nuclear burning may equally be important in accretion following the merger of a
He star with a NS or BH (\citealt{Fryer&Woosley98}).  \citet{Thone+11} proposed
that a He star-NS merger was responsible for the highly unusual ``Christmas''
gamma-ray burst 101225A, which exhibited exceptionally long gamma-ray and
thermal X-ray emission.  In order to explain the peak luminosity of the optical
`bump' following this event as supernova-like emission, the ejected mass of
$^{56}$Ni must have been small $\lesssim 0.1M_{\odot}$ (if one adopts the
distance advocated by \citealt{Thone+11}).  Future work on NuDAFs
will better address the accretion efficiency and $^{56}$Ni ejected due to
rapid He accretion, thereby allowing us to assess whether the high energy and
thermal optical emission from GRB 101225A was indeed compatible with a He
star-NS merger.

\section{Summary}
\label{s:summary}

This paper explores the effect of nuclear reactions on the evolution of
RIAFs, in the context of merging WDs and NSs or BHs. Two-dimensional
hydrodynamic simulations with FLASH3.2 are used to systematically characterize the
properties of these disks. Our main findings are the following:
\newline

\noindent 1. -- The effect of nuclear burning on RIAFs is controlled by the
	 	ratio $\Psi$ of the nuclear energy to the enthalpy at the radius $r_{\rm nuc}$
		where most of the fuel is consumed (eqns.~[\ref{eq:psi_def}] and [\ref{eq:rnuc_definition}]).
		The qualitative behavior of the system depends on the value of $\Psi$
		relative to a critical value $\Psi_{\rm crit}\sim 1$ which separates quiescent
		burning from large-scale detonations. The exact value of this critical parameter
		is sensitive to details about the disk, such as the thermal content or the
                amplitude of the viscous stress (Table~\ref{t:results}).
                \newline

\noindent 2. -- For disks that include cooling and/or have $\Psi \gtrsim \Psi_{\rm crit}$, a detonation
		can be triggered by hot spots formed near the burning front 
 		(Fig.~\ref{f:nudaf_ignition}). These
		temperature enhancements are generated by turbulent mixing of
		cold fuel and hot ash (Fig.~\ref{f:spot_profiles}), which produce an induction
		time-gradient consistent with that required to generate a detonation via the
		Zel'dovich mechanism (Fig.~\ref{f:spot_profiles}; Appendix \ref{s:zeldovich}).
		For low values of $\Psi$, detonations remain localized, and provide at most a
		slight enhancement to the mass ejection in quiescent disks. Increasing $\Psi$
		gives rise to detonations of increasing power.  
                \newline

\noindent 3. -- Exploding disks ($\Psi \gg \Psi_{\rm crit}$) can easily achieve expansion velocities 
		in excess of $1,000$~km~s$^{-1}$ (Table~\ref{t:results}). 
		A remnant disk, with a mass typically a few percent of the initial WD mass, 
		is left behind (Fig.~\ref{f:disk_mass_time}).
                \newline

\noindent 4. -- Non-exploding disks with nuclear reactions can generate unbound outflows
		along a funnel next to the rotation axis (Fig.~\ref{f:nudaf_individual}). 
		This material is composed primarily of nuclear ash.
		The expansion velocity of these outflows can also achieve $\sim 1000$~km~s$^{-1}$ 
		(Fig.~\ref{f:angular_quantities}), with mass outflow rates spanning the 
		range $10^{-5}$ -- $10^{-3}$ $M_\sun$~s$^{-1}$ (Table~\ref{t:results}).
                \newline

\noindent 5. --	The energy deposition by nuclear reactions locally enhances the turbulent
	  	kinetic energy, decreasing the mass accretion rate with increasing
		$\Psi$ (Fig.~\ref{f:mass_flux_machrms}). Given that a significant fraction
		of the outflowing material is bound and hence returns to the disk (Fig.~\ref{f:nudaf_individual}),
		this choking of accretion can prolong the lifetime of the disk well beyond a few
		viscous times. A significant amount of burnt material can then accumulate at large distances
		from the disk.
                \newline

\noindent 6. -- Outflows from the disk, generated either via quiescient disk
		winds or a large-scale detonation, may give rise to an optical supernova
		powered by the radioactive decay of $^{56}$Ni.  In the case of a disk
		detonation, some of the properties of the predicted transient are consistent
		with those of the observed class of unusual Type Ia SNe defined by SN 2002cx.
		\newline

This study has focused primarily on characterizing the range of outcomes
obtained in different regions of parameter space, and on identifying the main
parameter dependencies. Given the number of approximations made, our results
cannot be directly translated into observational predictions, nor do they
constitute definitive statements about the likely outcome of a realistic WD/NS
or WD/BH merger. Aside from including a realistic equation of state and a full
nuclear reaction network, a convergence study on the circum-torus medium needs
to be performed in order to reliably predict expansion velocities and
thermodynamic properties of the ejecta.  Also, fully three-dimensional
simulations are eventually desirable, since the formation and ignition of hot
spots is likely to occur at a position which is well-localized in azimuth, an
effect clearly not captured by an axisymmetric calculation. 

Even if NuDAFs in nature turn out not to explode, the fact that realistic
burning rates yield $\Psi > 1$ (Table~\ref{table:psivalues}) implies that the
structure and evolution of these disks will indeed be rich due to the multiple
elements to be burned and the dynamical importance of each of these reactions.
For example, despite the fact that model CO\_f1 [full
$^{12}$C($^{12}$C,$\gamma$)$^{24}$Mg rate] does not achieve explosion, the mass
loss rate in the unbound material is $\sim 100$ times the net accretion rate at
the inner boundary.  A significant fraction of the disk will
be ejected as an unbound flow at high velocities over a timescale $\sim
100t_{\rm orb} \sim$ few hr.  The observational signature of such an event may
constitute a unique type of transient, which is distinct from that produced in
the case of disk detonation or as yet observed.

A companion paper will continue the study of these disks using a realistic
equation of state and a full nuclear reaction network.

\acknowledgements

We thank Jim Stone, Stan Woosley, Lars Bildsten, Aristotle Socrates, 
James Guillochon,
Tobias Heinemann, Richard O'Shaughnessy, Chunglee Kim, Eliot Quataert, and Josiah Schwab 
for stimulating discussions. We also thank Hanno Rein for help with animations, and
Frank Timmes for making his stellar astrophysics subroutines publicly available.
An anonymous referee provided constructive comments that helped to improve the paper.
RF and BDM are supported by NASA through Einstein Postdoctoral Fellowship
grants number PF-00062 and PF-00065, respectively, awarded by the Chandra X-ray Center, 
which is operated by the Smithsonian Astrophysical Observatory for NASA under contract NAS8-03060.
The software used in this work was in part developed by the DOE NNSA-ASC OASCR Flash Center at the 
University of Chicago. Computations were performed at the IAS \emph{Aurora} cluster.

\appendix

\section{Unit conversion and burning rates}
\label{s:units_reactions}

For completeness, this Appendix show explicitly the conversion from physical to code units
and the functional forms of the reactions used.

The code takes the circularization radius $R_0$, the orbital velocity at this radius
$(GM_{\rm c}/R_0)^{1/2}$, and the maximum initial torus density
\begin{equation}
\label{eq:rho_unit}
\rho_{\rm uni} = 1.5\times 10^{5}\left(\frac{\rho_{\rm max}}{\bar\rho}\right)\left( \frac{M_{\rm WD}}{0.6\,M_\sun}\right)
\left( \frac{10^{9.3}\textrm{ cm}}{R_0}\right)^3\textrm{ g cm}^{-3},
\end{equation}
as the basic set of units. In equation~(\ref{eq:rho_unit}), $(\rho_{\rm max}/\bar\rho)$ is the ratio of the
maximum to average density in the torus. This quantity is obtained by computing the total torus 
mass using the density profile in equation~(\ref{eq:density_distribution}), being a function of the adiabatic
index and distortion parameter only.
For $\gamma=5/3$ and $d=\{1.2, 1.5,3\}$, the result is $\rho_{\rm max}/\bar\rho = \left\{0.536, 0.176, 2.47\times 10^{-2}\right\}$.
The temperature corresponding to $p/\rho = GM_{\rm c}/R_0$ is
\begin{equation}
\label{eq:T_unit}
T_{\rm uni} = 2\times 10^9 \left(\frac{\mu}{1.75} \right)\left( \frac{M_{\rm c}}{1.4M_\sun}\right)
\left( \frac{10^{9.3}\textrm{ cm}}{R_0}\right)\textrm{ K},
\end{equation}
where $\mu$ is given by equation~(\ref{eq:mmw}). The mean molecular weight for temperature conversion
purposes is calculated using the initial composition of the disk, once per model.
Given the pressure and density in code units, we use equations~(\ref{eq:rho_unit}) and (\ref{eq:T_unit}) 
to obtain physical density and temperature

The specific energy generation rates in physical units are converted to code units
through division by
\begin{eqnarray}
\frac{1}{R_0}\left(\frac{GM_{\rm c}}{R_0}\right)^{3/2} & \simeq &1.4\times 10^{16}
\left( \frac{M_{\rm c}}{1.4M_\sun}\right)^{3/2}\nonumber\\
& & \times\left( \frac{10^{9.3}\textrm{ cm}}{R_0}\right)^{5/2}\textrm{ erg g}^{-1}\textrm{ s}^{-1}.
\end{eqnarray}
The rate of change of mass fractions is converted to code units through multiplication
by 
\begin{equation}
\left(\frac{R_0^3}{GM_{\rm c}}\right)^{1/2} \simeq 6.5 \left( \frac{R_0}{10^{9.3}\textrm{ cm}}\right)^{3/2}
                                               \left( \frac{1.4M_\sun}{M_{\rm c}}\right)^{1/2}\textrm{ s}. 
\end{equation}

Analytic nuclear burning rates are taken from 
\citet{caughlan1988}\footnote{\url{http://www.phy.ornl.gov/astrophysics/data/cf88/}}.
The bulk of our study makes use of the $^{12}$C($^{12}$C,$\gamma$)$^{24}$Mg reaction, which
is the most energetic among the $\alpha$ reactions involving carbon, oxygen, and helium.
The specific energy generation rate is
\begin{eqnarray}
\label{eq:C12_full}
\dot{Q}_{\rm nuc,\,c12} &=&  3.96\times 10^{43}\rho_1 X_{\rm C}^2\,\frac{T_{A9}^{5/6}}{T_9^{3/2}}\times\nonumber\\
&&\exp{\left(-\frac{84.165}{T_{A9}^{1/3}}-2.12\times10^{-3}T_9^3\right)}\textrm{erg g}^{-1}\textrm{ s}^{-1},\nonumber\\
\end{eqnarray}
where $\rho_1$ is the density in g~cm$^{-3}$, $T_9 = T/(10^9\textrm{ K})$, $X_{\rm C}$ is the mass
fraction of $^{12}$C, and
\begin{equation}
T_{A9} = \frac{T_9}{1 + 0.0396T_9}.
\end{equation}
The rate of change of $X_{\rm C}$ is given by
\begin{equation}
\label{eq:C12_xdot}
\dot{X}_{\rm C,c12} = -\frac{m_{\rm C} }{Q_{12}}\,\dot{Q}_{\rm nuc,\,12},
\end{equation}
where $m_{\rm C} = 12m_n$ is the mass of a carbon nucleus, and $Q_{12} = 13.933$~MeV is the
energy liberated in the reaction.

To gain an analytic understanding of the effect of nuclear burning, we also use
an approximate power-law form of the $^{12}$C($^{12}$C,$\gamma$)$^{24}$Mg reaction.
Following a standard procedure (e.g., \citealt{Kippenhahn&Weigert94}), we expand equation~(\ref{eq:C12_full})
in a Taylor series in temperature around the point where nuclear burning is equal
to the dynamical time, and obtain
\begin{equation}
\label{eq:C12_pl}
\dot{Q}_{\rm nuc,\,pl} = 3.06\times 10^6 X_{\rm C}^2\,\rho_1\, T_9^{29}\,\textrm{erg g}^{-1}\textrm{ s}^{-1}. 
\end{equation}
The change in the carbon mass fraction is obtained from equation~(\ref{eq:C12_xdot}) by replacing
the energy generation rate. Good agreement between the two formulations is found in the temperature
range $[0.6,1.2]\times 10^9$~K. At higher temperatures, the full rate has an increasingly
weaker dependence on temperature relative to the power-law approximation.

To consider the case of helium white dwarfs, we also include the triple-alpha reaction.
The specific energy generation rate is
\begin{eqnarray}
\dot{Q}_{\rm nuc,\,3\alpha} & = & 5.07\times 10^8\,X_{\rm He}^3\,\rho_1^2\,T_9^{-3}\,
\exp{\left(-\frac{4.4027}{T_9}\right)} \nonumber\\
&& + 2.45\times 10^9\, r_{28} \,X_{\rm He}^3\,\rho_1^2\,T_9^{-3/2}
\exp{\left(-\frac{24.811}{T_9}\right)},\nonumber\\
\end{eqnarray}
where $X_{\rm He}$ is the mass fraction in helium nuclei. The coefficient $r_{28}$
is set to $1/10$. The rate of change of the helium mass fraction is 
\begin{equation}
\dot{X}_{\rm He} = -\frac{m_{\rm C}}{Q_{3\alpha}}\dot{Q}_{\rm nuc,\,3\alpha},
\end{equation}
where $Q_{3\alpha} = 7.275$~MeV is the energy released in the reaction.

\section{Verification Tests}
\label{s:code_tests}

This Appendix describes a series of tests conducted on the implementation
of the NuDAF setup in FLASH3.2 (\S\ref{s:time_dependent}).

We first tested the degree to which the torus can remain in steady-state in the
absence of viscous source terms. 
To this end, we use a torus with $d=1.125$, and an isothermal atmosphere
with $\rho_{\rm at} = 10^{-6}\rho_{\rm max}$ at $r_{\rm at} = 0.4R_0$ (eq.~[\ref{eq:isothermal_atmosphere}]).
The computational domain extends from $r_{\rm in} = 0.4R_0$ to $r_{\rm out} = 4R_0$, and
the resolution is $N_r = 64$, with 90 uniformly spaced cells covering the full range of polar angles.
Numerical diffusion 
causes a very small amount of mass to peel off from the torus edges and accrete
through the inner boundary. At our baseline resolution (\S\ref{s:time_dependent}), 
the total angular momentum in the computational domain is conserved to within a 
few parts in $10^{-6}$ over 10 orbits. The mass and angular momentum inside
torii (defined as the material inside an iso-density surface 
at $10^{-3}\rho_{\rm max}$), are conserved to within a few parts in $10^{-4}$ over 10 orbits.
Over the same time interval, the maximum torus density decreases by slightly more than $1\%$.

The inner boundary condition (eq.~[\ref{eq:boundary_conditions}]) was tested by 
evolving a torus similar to that used in the previous test,
but now with neither angular momentum
nor viscosity. Under such conditions, the system proceeds to free-fall towards the
gravitating mass. For an inner boundary with diameter comparable to the torus 
thickness $(r_{\rm in}=0.4R_0)$, the torus material is cleanly absorbed by 
the inner radial boundary, with no transients being generated. Using a smaller inner
radius $(r_{\rm in}=0.04R_0)$ causes some material to collide with the symmetry axis,
but the bulk of the torus is still cleanly accreted, with no discernible feedback from
the boundary.

\begin{figure}
\includegraphics*[width=\columnwidth]{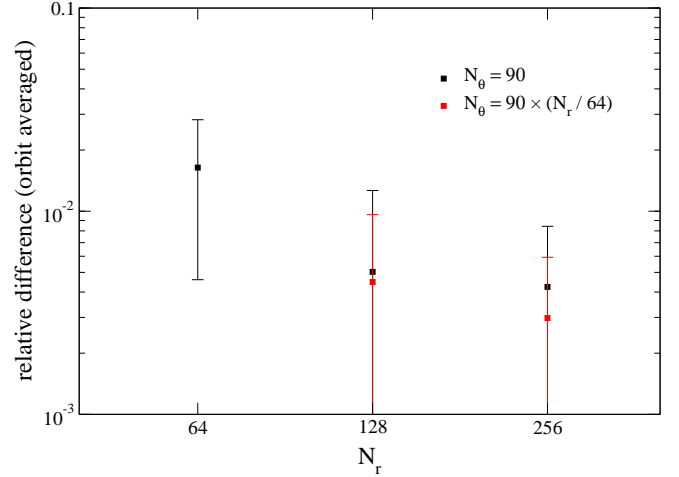}
\caption{Relative difference between the two terms appearing in equation~(\ref{eq:angular_momentum_test}),
normalized relative to the second term, and averaged between $20\%$ and $80\%$ of the first orbit. The torus
has $d=1.125$, inner and outer radii $r_{\rm in}=0.4R_0$ and $r_{\rm out}=4R_0$, respectively,
and a viscosity given by equation~(\ref{eq:viscosity_spb}) with $\tilde\nu_0 = 0.01$. The quantities
$N_r$ and $N_\theta$ denote the number of grid cells per decade in radius and in the polar direction,
respectively. The error bars denote the root-mean-square fluctuation of the difference, which is
due to fluctuations in the time derivative of equation~(\ref{eq:L2}), computed as a time-centered
finite difference with time step equal to $1/1000$ of an orbital period.}
\label{f:angz_cons_test}
\end{figure}

To test the accuracy of our viscous diffusion operator, we multiply equation (\ref{eq:angular_conservation})
by $\ell_z$, make use of equations (\ref{eq:mass_conservation}) and (\ref{eq:flux_divergence}), 
and integrate over volume, to obtain
\begin{equation}
\label{eq:angular_momentum_test}
\frac{\partial}{\partial t}L_2 + \int \totd^3 x\, \mathbf{F}_\ell \cdot \nabla \ell_z = 0,
\end{equation}
where
\begin{equation} 
\label{eq:L2}
L_2 = \int \totd^3 x \left(\frac{1}{2}\rho\ell_z^2\right),
\end{equation}
and the integral is carried out over the full computational domain. For 
equation~(\ref{eq:angular_momentum_test}) to be satisfied, the advection of $\ell_z$
and the flux formulation of the viscous operator must be treated correctly
(T. Heinemann, private communication).
Figure~\ref{f:angz_cons_test} shows the relative difference between the two terms in equation~(\ref{eq:angular_momentum_test}),
averaged over the first orbit, for a test model evolved with different spatial resolutions. 
The torus has $d=1.125$ and a viscosity proportional to density (eq.~[\ref{eq:viscosity_spb}]), with $\tilde{\nu_0} = 0.01$.
Values lie between $10^{-3}$ and $10^{-2}$ depending on resolution. A number of operations
are involved in computing this equation, which degrade its accuracy, thus we consider the results
as indicative of our correct treatment of viscous diffusion within the limitations of the hydrodynamic
method.

As an additional test of angular momentum transport, we compared the result of
using the flux-conservative formulation for angular momentum transport (eqns.~[\ref{eq:flux_divergence}] 
and [\ref{eq:angz_flux}]) with the result of directly computing the divergence of the viscous stress
tensor 
\begin{equation}
\label{eq:divT_phi}
(\nabla\cdot\mathbb{T})_\phi = \frac{1}{r^3}\frac{\partial}{\partial r}\left (r^3\,T_{\rm r\phi} \right) +
                       \frac{1}{\sin^2\theta}\frac{\partial}{\partial\theta}\left(\sin^2\theta\,T_{\rm \theta\phi} \right)
\end{equation}
as a 3-point finite difference operator. The result differs at the edges and surroundings of the
torus, with the flux formulation maintaining
a sharper torus surface (in $\ell_z$) for a longer time. Inside the torus, the evolution is nearly identical within
stochastic fluctuations induced by convection.

\section{Conditions for Detonation via Turbulent Mixing}
\label{s:zeldovich}

In this Appendix we evaluate whether  
the turbulent mixing of ash and fuel can generate the conditions for
a detonation in NuDAFs.
The Zel'dovich criterion for spontaneous initiation of a detonation
is that disparate portions of the burning region be separated by a distance
such that the difference in the timescale for nuclear heating implies a
supersonic phase velocity (\citealt{Zeldovich+70};
\citealt{Blinnikov&Khokhlov87}; \citealt{Woosley90}).  Quantitatively, this
condition is expressed as (e.g.,~\citealt{Niemeyer&Woosley97};
\citealt{bell2004})
\be
\left|\nabla t_{\rm ind}\right|^{-1} > c_s,
\label{eq:zeld}
\ee
where $c_s$ is the adiabatic sound speed and $t_{\rm ind}$ is the induction
time, which is loosely defined as the time required to `run away' to high
temperatures via burning at constant pressure.  Since $t_{\rm ind} \propto
c_{s}^{2}/\dot{Q}_{\rm nuc}$ is a rapidly decreasing function of temperature, a
shallow temperature gradient is required across the burning region for
triggering a  detonation.  Such a condition may be satisfied in NuDAFs due to
the effects of turbulent mixing. 

We focus on the burning of fuel with an initial mass fraction $X_{\rm f0}$.
The steady-state structure of the accretion flow is characterized by two
radially-separated regions: (1) unburned fuel with temperature $T_{\rm f}$ and
density $\rho_{\rm f}$; and (2) burned ash with temperature $T_{\rm a}$ and
density $\rho_{\rm a}$.  The burning front separating the upstream (region 1)
from the downstream (region 2) is centered about the radius $r \equiv r_{\rm
nuc}$ (eq.~[\ref{eq:rnuc_definition}]) and has a width $\Delta r_{\rm nuc}$.
The radius $r_{\rm nuc}$ is approximately determined by equality of the fuel
consumption time $t_{\rm nuc} = X_{\rm f}/\dot{X_{\rm f}}$ and the dynamical
timescale $t_{\rm dyn} = (r^{3}/GM)^{1/2}$.

If burning occurs at constant pressure, then conservation of mass and enthalpy
determines the change in density and temperature across the burning front
(e.g.,~\citealt{Khokhlov+97}):
\begin{eqnarray}
\frac{\rho_{\rm a}}{\rho_{\rm f}} = \frac{T_{\rm f}}{T_{\rm a}} = \frac{1}{1 + \Psi}\,\,\,\,\,\,\,\,\,\,\,&(P_{\rm gas} \gg P_{\rm rad})&, \\
\frac{\rho_{\rm a}}{\rho_{\rm f}} = \frac{1}{1 + \Psi};\,\,\,\,T_{\rm f} = T_{\rm a}\,\,\,&(P_{\rm rad} \gg P_{\rm gas}),&
\end{eqnarray}
where 
\be
\Psi = (\gamma - 1)\frac{\varepsilon_{\rm nuc}}{c_{\rm s,f}^{2}}
\label{eq:Q}
\ee
is the ratio of the specific nuclear energy released per reaction (eq.~[\ref{eq:enuc_definition}])
to the enthalpy of the fuel $c^2_{\rm s,f}/(\gamma-1)$ (see also equation~[\ref{eq:psi_def}]).
We have separated cases corresponding to
whether gas or radiation pressure dominates.  

Due to the turbulent nature of accretion and RT instabilities, the burning
front at $r = r_{\rm nuc}$ is not completely smooth
(e.g., Fig.~\ref{f:nudaf_ignition}).  Instead, 
parcels of burned ash occasionally mix
with the colder fuel upstream.  If mixing produces conditions such that
equation (\ref{eq:zeld}) is satisfied over the length of the eddy $L_{\rm
edd}$, then a detonation may be triggered.  If a fraction $f_{\rm a}$ of ash is
mixed with a fraction $1-f_{\rm a}$ of fuel (resulting in a reactant mass fraction
$X_m = [1-f_{\rm a}]X_{\rm f0}$), then the density $\rho_{\rm m}$ and temperature $T_{\rm m}$ of
the resulting mixture are also determined by conservation of mass and enthalpy:
\begin{eqnarray}
\frac{\rho_{\rm m}}{\rho_{\rm f}} = \frac{T_{\rm f}}{T_{\rm m}} = \frac{1}{1 + f_{\rm a}\Psi}\,\,\,\,\,\,\,\,\,\,\,&(P_{\rm gas} \gg P_{\rm rad})&, \\
\frac{\rho_{\rm m}}{\rho_{\rm f}} = \frac{1}{1 + f_{\rm a}\Psi};\,\,\,\,T_{\rm m} = T_{\rm f}\,\,\,&(P_{\rm rad} \gg P_{\rm gas}),&
\label{eq:mixed}
\end{eqnarray}

The length of an eddy $L_{\rm edd}$ which is capable of efficient mixing must
obey $\Delta r_{\rm nuc} \lesssim L_{\rm edd} \lesssim H = r_{\rm nuc}/2$ since
it must both fit inside the disk midplane and it must sample both fuel and ash
over the width of the mixing region $\Delta r_{\rm nuc}$.  The condition for
detonation in equation (\ref{eq:zeld}) can thus approximately be written as
\be 
\left|\nabla t_{\rm ind}\right|^{-1} \sim \frac{L_{\rm edd}}{t_{\rm ind}(X_{\rm m},\rho_{\rm m},T_{\rm m})} > c_{s}
\ee
or
\be
\left(\frac{L_{\rm edd}}{H}\right) > \frac{t_{\rm ind}(X_{\rm m},\rho_{\rm m},T_{\rm m})}{t_{\rm orb}(r_{\rm nuc})},
\label{eq:zeld2}
\ee
where we have assumed that the sound speed of the mixture is similar to that in
the disk midplane, and we have used the condition of vertical hydrostatic
equilibrium $c_{s} = H/t_{\rm orb}$.  

Assuming a burning rate of the general form $\dot{Q}_{\rm nuc} \propto
\varepsilon_{\rm nuc}\,X^{q}\rho^{q-1}T^{\beta}$ (eq.~[\ref{eq:C12_pl}]), one has that
$t_{\rm ind} \propto \rho^{-(q-1)}X^{-q}T^{-\beta}/[\Psi\,\bar{\beta}]$ for burning at
constant pressure, where the 
factor of $\bar{\beta}$ in the denominator accounts for the fact that runaway to high
temperatures occurs on a timescale which is shorter than the heating timescale
estimated using the initial heating rate when gas pressure dominates 
($\bar{\beta} \simeq \beta$ when $P_{\rm gas} \gg P_{\rm rad}$; 
$\bar{\beta} \simeq 1$ when $P_{\rm rad} \gg P_{\rm gas}$).  Since $t_{\rm nuc} = X/\dot{X}
\propto \rho^{-(q-1)}X^{-(q-1)}T^{-\beta}$, we also have
$t_{\rm nuc} \simeq \Psi \bar{\beta} t_{\rm ind}$.  Using the relation
$t_{\rm nuc}(X_{\rm f},\rho_{\rm f},T_{\rm f}) = t_{\rm dyn}$ that determines
$r_{\rm nuc}$, one can rewrite equation (\ref{eq:zeld2}) as
\be
\frac{L_{\rm edd}}{H} > \frac{1}{2\pi \bar{\beta} \Psi}\left[\frac{t_{\rm ind}(X_{\rm m},\rho_{\rm m},T_{\rm m})}{t_{\rm ind}(X_{\rm f},\rho_{\rm f},T_{\rm f})}\right],
\ee
or
\begin{eqnarray}
\frac{L_{\rm edd}}{H} & > &\frac{1}{2\pi\bar{\beta} \Psi}\left(\frac{c_{\rm s,m}^{2}}{c_{\rm s,f}^{2}}\right)\left(\frac{\rho_{\rm f}}{\rho_{\rm m}}\right)^{q-1}\left(\frac{X_{\rm f}}{X_{\rm m}}\right)^{q}\left(\frac{T_{\rm f}}{T_{\rm m}}\right)^{\beta}\nonumber \\
 & > &  \frac{1}{2\pi \Psi}\left(\frac{1}{1-f_{\rm a}}\right)^{q} 
\left\{
\begin{array}{lr}
  \beta^{-1}\left(1+f_{\rm a}\Psi\right)^{q-\beta}
, &
\quad(P_{\rm gas}/P_{\rm rad}) \gg 1 \\
\noalign{\smallskip}
 (1+f_{\rm a}\Psi)^{q}
, &
\quad(P_{\rm rad}/P_{\rm gas}) \gg 1 \\
\end{array}
\right., \nonumber \\
\label{eq:zeldcond}
\end{eqnarray}
where we have used the results from equation (\ref{eq:mixed}).

Equation (\ref{eq:zeldcond}) shows that in a gas pressure-dominated flow, even
small eddies with $L_{\rm edd} \sim 0.1 H$, such as those producing detonations
in our simulations (Fig.~\ref{f:nudaf_ignition}), require only a moderate
amount of entrained fuel $f_{a} \ll 1$ to satisfy the condition for a
detonation given a temperature-sensitive reaction (e.g., $\beta = 29$, $q = 2$;
eq.~[\ref{eq:C12_pl}]).  The mixed fraction of upstream ash is indeed small
($X_{a} \sim$ few $\%$) in the regions which form hot spots and detonations in
our simulations (see Fig.~\ref{f:spot_profiles}).  The sensitive dependence of
the bracketed quantity on the RHS on $\Psi$ (almost exponential) may in part
also explain why the formation of a detonation in our simulations depends
sensitively on this value.
On the other hand, in a radiation pressure dominated disk the condition
(\ref{eq:zeldcond}) is more challenging to satisfy, 
as the lack of temperature fluctuations cause mixing to be detrimental given
the decrease in density (eq.~[\ref{eq:mixed}]).

Producing a detonation is 
harder than
satisfying condition
(\ref{eq:zeldcond}), since 
mixing must occur before
complete burning.  Another important parameter is thus the ratio of the
characteristic eddy turnover timescale 
$t_{\rm edd} \sim L_{\rm edd}/v_{\rm edd} = \mathcal{M}^{-1}\,L_{\rm edd}/c_s$, which sets the timescale for
mixing, to the burning timescale of the mixture $t_{\rm nuc}:$ 
\begin{equation}
\label{eq:eddburn}
\frac{t_{\rm nuc}}{t_{\rm edd}} = \mathcal{M}\Psi \bar{\beta}\left(\frac{c_s}{L_{\rm edd}/t_{\rm ind}}\right),
\end{equation}
where $\mathcal{M} \equiv v_{\rm edd}/c_{s} < 1$ is the turbulent Mach number. 
Provided that the condition for detonation (eq.~[\ref{eq:zeld2}])
is marginally achieved, equation (\ref{eq:eddburn}) shows that 
efficient mixing requires a combination of vigorous turbulence and/or strong energy release.
Dominance of radiation pressure makes mixing more inefficient, which is helpful
in satisfying equation~(\ref{eq:zeldcond}), as a higher mixing fraction results in a larger required
eddy size.

\bibliographystyle{apj}
\bibliography{nudaf,apj-jour}

\end{document}